\documentclass[structabstract]{aa}
\usepackage{multirow}
\usepackage{graphicx}
\usepackage{rotating}
\usepackage{subfigure}
\usepackage{color}
\usepackage{txfonts}
\usepackage{natbib}
\bibpunct{(}{)}{;}{a}{}{,}                     

\begin{document}
   \title{Dielectronic Recombination of Argon-Like Ions}

   \author{D.~Nikoli\'c$^1$, T.~W.~Gorczyca$^1$, K.~T.~Korista$^1$, and N.~R.~Badnell$^2$}

   \institute{
   $^1$Department of Physics, Western Michigan University, Kalamazoo, MI 49008, USA\\
   $^2$Department of Physics, University of Strathclyde, Glasgow G4 0NG, UK\\
   \email{gorczyca@wmich.edu}
   }

   \date{}


  \abstract
   {
   We present a theoretical investigation of dielectronic recombination (DR) of Ar-like ions that sheds new light
   on the behavior of the rate coefficient at low-temperatures where these ions form in photoionized plasmas.
   }
   {
   We provide results for the total and partial Maxwellian-averaged DR rate coefficients from the
   initial ground level of \ion{K}{ii}--\ion{Zn}{xiii}  ions. It is expected that these new results will
   advance the accuracy of the ionization balance for Ar-like M-shell ions and pave the way towards a
   detailed modeling of astrophysically relevant X-ray absorption features.
   }
   {
   We utilize the AUTOSTRUCTURE computer code to obtain the accurate core-excitation thresholds in target
   ions and carry out multiconfiguration Breit-Pauli (MCBP) calculations of the DR cross section in the independent-processes,
   isolated-resonance, distorted-wave (IPIRDW) approximation.
   }
   {
   Our results mediate the complete absence of direct DR calculations for certain Ar-like ions and question
   the reliability of the existing empirical rate formulas, often inferred from renormalized data within this
   isoelectronic sequence.
   }
   {}

   \authorrunning{D.~Nikoli\'c {\it et al}}
   \maketitle

\section{Introduction}
\label{Sec:Intro}
Atomic structure and dynamic behavior of highly-charged ions is one of the key ingredients
presently required for both laboratory plasma diagnostics and interpretation of astrophysical
phenomena \citep{Fawcett:1991,Liedhal:2000,Kallman:2007}.
The most common diagnostic technique to probe for electron temperature of laser-produced,
fusion, or astrophysical plasmas involves spectroscopical observations of intensity ratios of
EUV or X-ray emission lines coming from consecutive ionization stages of a single plasma
component. The accuracy of derived plasma parameters is strongly affected by uncertainties
in chemical abundances, often inherited through the use of unreliable collisional ionization
and/or dielectronic recombination (DR) rates \citep{Savin:2002}.
\cite{Seon:2003-Ti-Cr} investigated the effect of the uncertainties in DR rates on
an isoelectronic line ratio in \ion{Ti}{} and \ion{Cr}{} plasmas and found substantial
differences in the fractional abundances obtained for \ion{Ti}{v} and \ion{Cr}{vii} ions
(a shift of the curves to lower temperatures) as compared to those inferred using the recommended
recombination rate coefficients \citep{Mazzotta:1998}.

In recent years an enormous amount of progress, both theoretically and experimentally, has been
made in improving the DR rate database along isoelectronic series within the first and second rows,
and the third row up through Mg-like ions (see the review by \cite{Kallman:2007}, and references therein).
In most cases it is found that the newly determined DR rates are significantly larger than their
earlier recommended estimates, having profound consequences on the ionization balance and thermal
equilibrium in both photoionized and collisionally ionized plasmas, from the solar corona to
Active Galactic Nuclei (AGN) (see, for example, \cite{Chakravorty:2008,Chakravorty:2009,Bryans:2006,Bryans:2009:a,Bryans:2009:b,Dere:2009}).
In this work, we present improved theoretical predictions of the DR rates of Ar-like ions. Of these ions,
those of the iron peak elements are some of the more abundant in cosmic plasmas. In much of the
remainder of this introduction we will provide some of the motivation in improving their atomic
database, in particular their DR rate coefficients.

The strong contribution of M-shell \ion{Fe}{} ions to the unresolved transition array of inner-shell
absorption lines in $\sim$15-17~{\AA} X-ray spectra of several AGN observed with XMM-Newton and
Chandra, was initially not well understood \citep{Netzer:2001}. The fact that AGN photoionization
models initially overpredicted the average ionization stage of iron was attributed in part to an
underestimate in the low-temperature DR rate coefficients for M-shell iron \citep{Ferland:2004}, and constituted
the main motivation behind the benchmark calculations recently performed by \cite{Badnell:Fe3pq:2006},
as well as experimental and theoretical results presented in \cite{Lukic:2007}.
The present work further extends the calculations of \cite{Badnell:Fe3pq:2006} for \ion{Fe}{ix} ions
by augmenting the configuration interaction (CI) with some of the most important
$\Delta n_{{\rm c}} = 0$ ionic core excitations \citep{Aggarwal:2006, Zeng:2006}.

In the framework of testing nucleosynthesis models, \cite{Ellison:2001} identified \ion{Co}{}
as a rewarding element to study galactic and stellar formation histories through the observed
abundance trends \citep{Peloso:2005}.
For example, the agreement of the modeled time-dependent ejecta compositions and velocities
with observed Type 1a supernovae spectra requires a substantial initial presence of \ion{Ni}{},
\ion{Co}{}, and \ion{Fe}{} in the outer layers of the ejecta \citep{Hillebrandt:2000}.
In the present work, we study the photorecombination of \ion{Co}{x} as an example
of a heavier iron peak element for which the stellar photospheric elemental abundances are
less well known \citep{Adelman:2000}.

Nickel is one of the most important heavy impurities in tokamaks and early attempts by the
\cite{TFR:1980} to model the fractional abundance of its charge states suffered from
deficient DR data. Recent simulations of the observed plasma emission
from magnetic confinement fusion devices, namely the JET tokamak in Abingdon and RFX in
Padova \citep{Mattioli:2004}, also lacked accurate recombination rate coefficients for the \ion{Ni}{xi} ion.
However, reliable electron impact ionization data of the remaining ions in the nickel isonuclear
sequence have been provided by \cite{Pindzola:1991} and were widely used by \cite{Mattioli:2004}
in simulations of Ni emission line spectra.
The past few years have marked a renewed theoretical interest \citep{Verma:2007,Aggarwal:2007,Aggarwal:2008}
in electron excitation data for argon-like nickel, initiated by its identification in numerous
astrophysical plasmas (consult \cite{Verma:2007} for an exhaustive up-to-date bibliography).
The most recent studies of the Intra-Cluster Medium, as discussed by \cite{Werner:2008},
put constraints on supernova models by using Ni/Fe abundance patterns in the ejecta of type Ia supernovae.
In addition, from the Mass Time-of-Flight Spectrometer data accumulated during the first decade of SOHO's
operation, \cite{Karrer:2007} inferred charge-state distributions, isotopic composition, and the
elemental \ion{Ni}{}/\ion{Fe}{} ratio of the solar wind, and confirmed that both nickel and iron become
enriched in the solar corona.

The significance of reliable atomic data has been demonstrated by \cite{Churazov:2004} through the
modeling and interpretation of the 5-9~keV spectrum from the multi-temperature core of the Perseus galaxy cluster.
In their study, \cite{Churazov:2004} used APEC \citep{Brickhouse:2001} and MEKAL \citep{Mewe:1985}
models, both having the redshift of major line energies, the heavy element abundances, and the plasma temperatures as
free parameters. It has been shown that the MEKAL model yields the best description of the spectra only when nickel
is overabundant relative to iron by a factor of $\sim 2$ compared to solar. However, this enhancement is not
required by the simulations of the APEC code ({http://cxc.harvard.edu/atomdb})
that uses updated atomic data.

The present computational study is part of an ongoing investigation of DR processes in argon-like
ions \citep{Nikolic:2007, Nikolic:2009, Nikolic:2010}, and deals with $\Delta n_{{\rm c}} = 0,\,1$ ionic core excitations and
associated dielectronic resonances that dominate electron-ion recombination in photoionized plasmas.
The theoretical foundation and computational method we use for the DR calculations are found elsewhere
\citep{Badnell:2003} and here we only outline the essence.
With the use of the open-source AUTOSTRUCTURE code \citep{Badnell:1986,Badnell:1997}, we carry out MCBP
computations of energy levels and decay rates in an intermediate coupling scheme for
Ar-like \ion{K}{ii}, \ion{Ca}{iii}, \ion{Sc}{iv}, \ion{Ti}{v}, \ion{V}{vi}, \ion{Cr}{vii}, \ion{Mn}{viii},
\ion{Fe}{ix}, \ion{Co}{x}, \ion{Ni}{xi}, \ion{Cu}{xii}, and \ion{Zn}{xiii} ions.
In order to account efficiently for all DR contributions coming from numerous Rydberg series of
resonances and offer them to the plasma modeling community in convenient level-resolved format,
we further enforce the independent-processes, isolated-resonance, distorted-wave (IPIRDW) approximation \citep{Pindzola:1991:IPIR}.

In the remainder of this paper, we will proceed as follows. Section~\ref{Sec:eproc} discusses the process of
electron-ion recombination and transparently outlines the main relations and equations arising within the adopted
methodology. A comparative overview of existing atomic structure in argon-like ions is provided throughout
Section~\ref{Sec:ASstructure}, and an analysis of the results is presented in Section~\ref{Sec:Analysis}.

\section{Elementary processes of relevance}
\label{Sec:eproc}
The contribution of the photorecombination process of an ionization
state $q+$ from a single partial wave $(J,\,\pi)$ can be described as
\begin{eqnarray}\label{Eq:process}
\left[{\rm{e}}^{-}+{\rm{I}}^{(q)+}(3s^{2}3p^{6})\right]_{J,\,\pi} & \to &
\hspace*{-2mm}{\rm{I}}^{(q-1)+\ast\ast}\left(
\begin{array}{l}
    3s^{2}3p^{5}\,3d n\ell \\
    3s3p^{6}\,3d n\ell \\
    3s^{2}3p^{5}\,4\ell' n\ell \\
    3s3p^{6}\,4\ell' n\ell
\end{array}\right)_{J,\,\pi}\nonumber \\[-3mm]
 & \raisebox{2mm}{$_{\rm{RR}}\searrow \raisebox{3mm}{$\;\;{}^{\rm{DR}}$}$} &
 \hspace*{-8mm}\begin{array}{l}\raisebox{2mm}{\hspace*{+7mm}$\downarrow$} \\ {\rm{I}}^{(q-1)+\ast}(3s^{2}3p^{6}\,n\ell) + \omega \;.\end{array}
\end{eqnarray}
Here $c$ is a continuum state consisting of an initial electron incident upon the target ion
${\rm{I}}^{(q)+}$ that is either directly captured, via radiative recombination (RR), to a
${\rm{I}}^{(q-1)+\ast}$ bound state, $b$, or captured into an autoionizing ${\rm{I}}^{(q-1)+\ast\ast}$
doubly-excited state, $d$, that undergoes subsequent radiative decay to the same final bound state, $b$,
completing the DR process.
The present work investigates $\{3s, 3p\} \to \{3d, 4\ell'\}$ inelastic excitations from
the ${\rm{I}}^{(q)+}$ ground state that give rise to Rydberg series of autoionizing
states formed by the capture of the scattered electron. Each Rydberg series of resonances will
converge to a corresponding threshold, given in Tables~\ref{Tab:K:Ca:tar-str}, \ref{Tab:V:Cr:tar-str}, \ref{Tab:Mn:Fe:tar-str}, \ref{Tab:Co:Ni:tar-str}, and \ref{Tab:Cu:Zn:tar-str}.

Within the adopted IPIRDW framework, our calculations rely on lowest-order perturbation
theory to compute Lorentzian resonance profiles as a function of the electron's center-of-mass
(c.m.) energy $\epsilon$. The total DR cross section is then given as
\begin{equation} \label{Eq:DR-IPIR}
\sigma^{{\rm DR}}(\epsilon) = \frac{1}{\epsilon} \frac{2}{\pi}
\sum_{J,\,\pi} \sum_{b} \sum_{d}^{N^{J,\,\pi}} \frac{ \mathcal{S}_{c \to b}^{d} / \overline\Gamma_{d}}{1+\varepsilon_{d}^{2}}\;.
\end{equation}
Here $\varepsilon_{d}=2(\epsilon - \epsilon_{d})/\overline\Gamma_{d}$ is the
reduced c.m. energy at which the (partial) integrated resonant strength is
\begin{equation} \label{Eq:Sint} \mathcal{S}_{c \to b}^{d} [\mbox{Mb Ry$^{2}$}] = 2.674\cdot10^{-14}\,
\frac{g_d}{2g_{ion}} \frac{A^{a}_{d \to c}\,A^{r}_{d \to b}}{\sum_{c'} A^{a}_{d \to c'} + \sum_s A^{r}_{d \to s}}\;,
\end{equation}
with $g_d$ and $g_{ion}$ as statistical weights of the resonant state $d$ in the recombined
ion and the ground state of the target ion, respectively. The summations over $c'$ and $s$
in Eq.~(\ref{Eq:Sint}) cover all states that are attainable from resonant state $d$
either by radiative decay or by autoionization, with corresponding rates given in inverse seconds.
Hence, the sum over $s$ includes not only bound states that are below the first ionization limit
of the recombined ion, $E_{th}^{(1)}$, but also may consider a radiative cascade through other
autoionizing states resulting in the total radiative rate, $A^{r}_{d}$.
The sum over $c'$ accounts for resonant scattering (excitation) and amounts to the total
autoionization rate, $A^{a}_{d}$. In Eqs.~(\ref{Eq:DR-IPIR})-(\ref{Eq:Sint}) we implicitly
assume that both the resonance position, $\epsilon_d = E_d - E_{th}^{(1)}$, and the total width,
$\overline\Gamma_{d} = \hbar (A^{a}_{d}+A^{r}_{d})$, are in Rydberg units.
In addition, the summation over $d$ spans all of the $N^{J,\,\pi}$ doubly excited states of
given parity $\pi$ and total angular momentum $J$ that are formed through Eq.~(\ref{Eq:process})
for $3\leq n \leq 1000$ and $0\leq \ell \leq 10$, wherein the index $b$ considers all accessible bound states.

The DR/RR rate coefficient (in units of ${\rm cm}^{3}\,{\rm s}^{-1}$) for ${\rm I}^{q+}$ ions in
a plasma with a Maxwellian electron energy distribution $f_{{\rm MB}}(\epsilon,\,T)$ is given by
\begin{equation} \label{Eq:MaxwRate}
\alpha^{{\rm DR/RR}}(T) = \int_{0}^{\infty}
v(\epsilon)\,\sigma^{{\rm DR/RR}}(\epsilon)\;f_{{\rm MB}}(\epsilon,\,T)\,d\epsilon
\end{equation}
and conveniently modeled using physically-motivated fitting formulae
\begin{equation}\label{Eq:Fit:DR}
 \alpha^{{\rm DR}}(T) = T^{-3/2} \sum_{i} c_{i} \exp(- E_{i} / T )\;\;{\rm and}
\end{equation}
\begin{equation}\label{Eq:Fit:RR}
 \alpha^{{\rm RR}}(T) = A \, \left[ \sqrt{T/T_{0}} \left( 1 + \sqrt{T/T_{0}} \right)^{p_{-}}
 \left( 1 + \sqrt{T/T_{1}} \right)^{p_{+}} \right]^{-1} \;,
\end{equation}
where the more general form $p_{\pm} =1 \pm B \pm C \exp( - T_{2} / T )$ is used here.
The fitted DR rate coefficients given by Eq.~(\ref{Eq:Fit:DR}) are relatively smooth and broad
curves with several local maxima, each at a particular temperature $T_i=2E_{i}/3$, with $E_i$
given in Table~\ref{Tab:DR:Maxwl}. On the other hand, the RR rate coefficients given by
Eq.~(\ref{Eq:Fit:RR}) decrease monotonically with temperature; the fitting coefficients
$A$, $B$, $C$, and $T_{0,1,2}$ are given in Table~\ref{Tab:RR:Maxwl}. These nonlinear
least-squares fits are accurate to better than 1\% over $(10^1 - 10^7)q^2$~K with the
correct asymptotic forms outside of this temperature range.

\section{Atomic structure}
\label{Sec:ASstructure}
Before performing the electron-ion scattering calculation in Eq.~(\ref{Eq:process}),
it is necessary to obtain an accurate description of the Ar-like target-ion states.
Our approach for the entire Ar-like isoelectronic sequence is essentially the
same as what was done earlier for \ion{Ti}{v} \citep{Nikolic:2009}.
Since the dominant DR contribution is due to the $e^{-}+3s^{2}3p^{6}\rightarrow 3s^{2}3p^{5}3d^{}\,n\ell$
capture, we are particularly concerned with obtaining highly-accurate wavefunctions for both the
$3s^{2}3p^{6}$ ground state and the $3s^{2}3p^{5}3d^{}$ excited state.  However, we also need to consider
the $3s^{}3p^{6}3d^{}$, $3s^{2}3p^{5}4\ell'$, and $3s^{}3p^{6}4\ell'$ target states that also contribute
to the DR rate coefficient via Eq.~(\ref{Eq:process}). Single and double promotions out of these configurations
are also included in our total configuration expansion for each target state.

With this configuration basis, the $1s$, $2s$, $2p$, $3s$, $3p$, and $3d$ Slater-type orbitals were determined
by varying the radial scaling parameters, $\lambda_{n\ell}$, to minimize the equally-weighted-sum of MCBP eigenenergies
of the seventeen lowest states, using the model potential of \cite{Burgess:89} with mass-velocity and Darwin corrections.
The values of the radial scaling parameters for closed-core orbitals ($1s$, $2s$, and $2p$) remained throughout the optimization
in near proximity to their default value of $1.0$.
We then applied small adjustments to the $3\ell$ scaling parameters in order to reproduce the NIST experimental
$3p\rightarrow 3d$ radiative data of \cite{Shirai:2000}.  The resultant $\lambda_{n\ell}$ values are listed
in Table~\ref{Tab:lambdas} and the  radiative data are listed in  Table~\ref{Tab:tar-rad}.
As is seen in Table~\ref{Tab:tar-rad}, our target description is such that the present oscillator strengths
are in excellent agreement with the experimental values.

We have also confirmed that the computed eigenenergies were stationary with respect to the small variations
in $\lambda_{n\ell}$ values that were used to fine-tune the $3p\rightarrow 3d$ oscillator strengths. Indeed,
as is seen in Tables~\ref{Tab:K:Ca:tar-str}, \ref{Tab:V:Cr:tar-str}, \ref{Tab:Mn:Fe:tar-str}, \ref{Tab:Co:Ni:tar-str},
and \ref{Tab:Cu:Zn:tar-str} for the lowest-lying Ar-like states, our computed energies are in excellent agreement
with the experimental values given in the NIST Atomic Spectra Database~\citep{Sugar:1985,Shirai:2000,Pettersen:2007}.
In the case of higher ionization stages, the atomic data produced by \cite{Irimia:2003} using the multi-configurational
Hartree-Fock (MCHF) method, or by a non-relativistic single-configuration approach of \cite{Ghosh:1997}, are up to
7.2\% lower than the NIST values. The basic CIV3 atomic data for \ion{Fe}{ix} produced by \cite{Verma:2006} are
noticeably closer to the present results than those of \cite{Aggarwal:2006} using the fully relativistic
multi-configurational Flexible Atomic Code (FAC) of \cite{FAC}.

\section{DR Results}
\label{Sec:Analysis}

\begin{figure*}[!hbtp]
\centering
\subfigure{\resizebox{0.48\textwidth}{!}{\includegraphics[angle=0,scale=1.0]{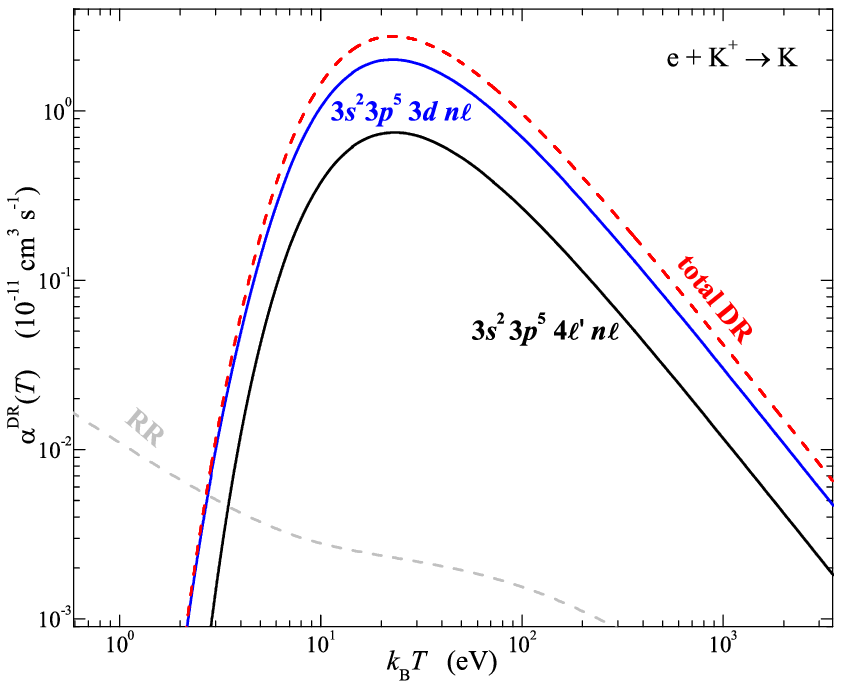} }\label{Fig:K:ASMaxwl}}
\subfigure{\resizebox{0.48\textwidth}{!}{\includegraphics[angle=0,scale=1.0]{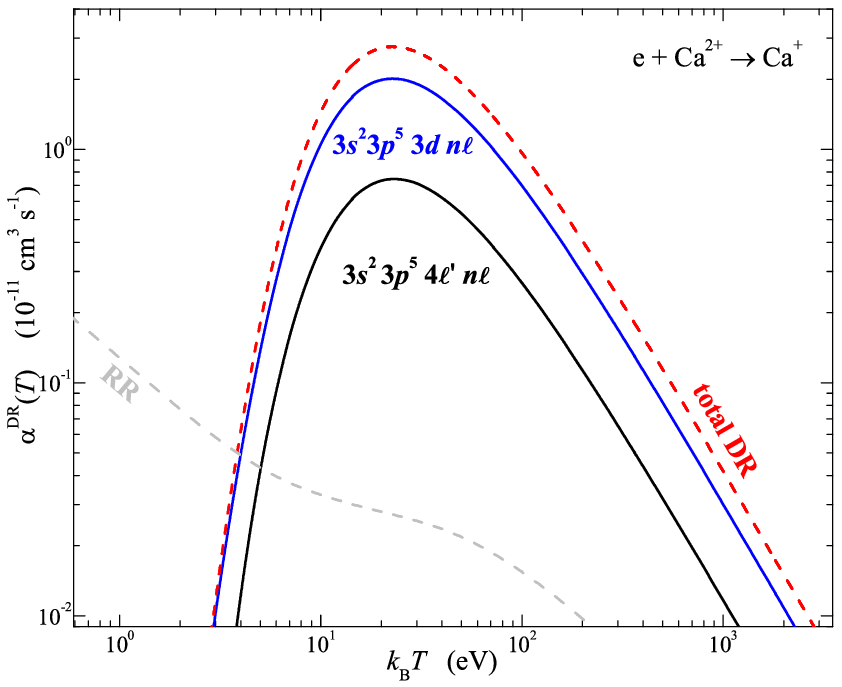} }\label{Fig:Ca:ASMaxwl}}
\caption[]{\label{Fig:K:Ca:ASMaxwl}
Present Maxwellian-averaged DR and RR rate coefficients for \ion{K}{ii} (left) and \ion{Ca}{iii} (right).
The red and gray dashed curves are the total DR and RR rate coefficients, respectfully. The intra-shell $\Delta n_{{\rm c}} = 0$
contributions (solid blue curve) are due to dominant $3p\to 3d$ core excitations; the inter-shell $\Delta n_{{\rm c}} = 1$ contributions
(solid black curve) are due to $3p\to 4\ell'$ promotions.
}
\end{figure*}

\begin{figure*}[!hbtp]
\centering
\subfigure{\resizebox{0.48\textwidth}{!}{\includegraphics[angle=0,scale=1.0]{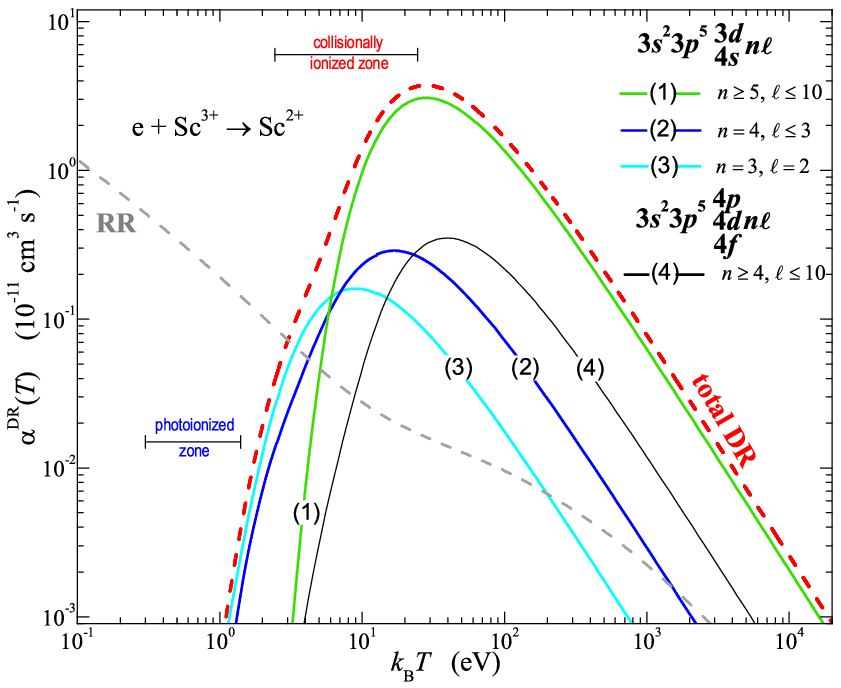} }\label{Fig:Sc:ASMaxwl}}
\subfigure{\resizebox{0.48\textwidth}{!}{\includegraphics[angle=0,scale=1.0]{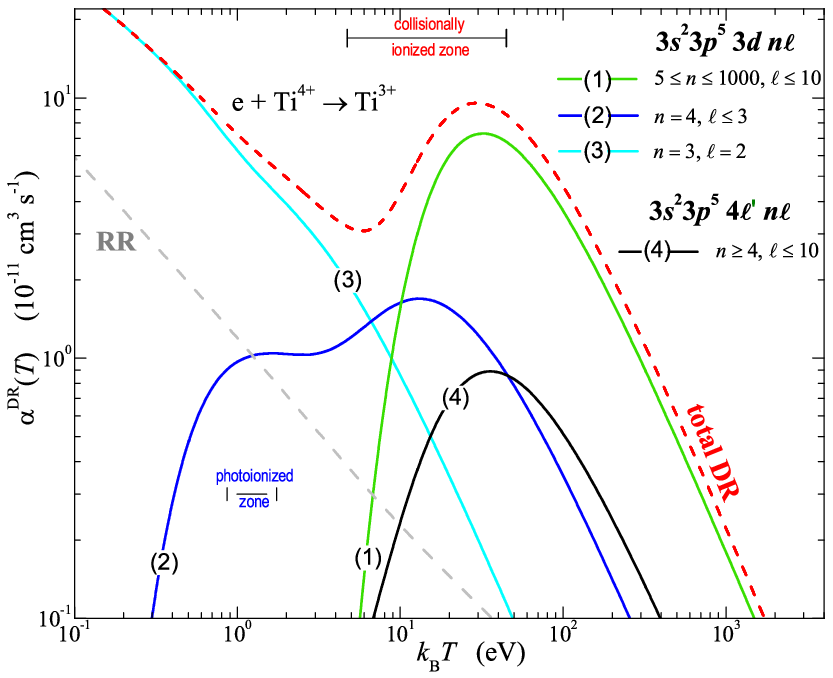} }\label{Fig:Ti:ASMaxwl}}
\caption[]{\label{Fig:Sc:Ti:ASMaxwl}
Present Maxwellian-averaged DR and RR rate coefficients for \ion{Sc}{iv} (left) and \ion{Ti}{v} (right).
The red and gray dashed curves are the total DR and RR rate coefficients, respectfully. The intra-shell $\Delta n_{{\rm c}} = 0$
contributions are due to dominant $3p\to 3d$ core excitations; the inter-shell $\Delta n_{{\rm c}} = 1$ contributions are due to $3p\to 4\ell'$ promotions.
}
\end{figure*}

\begin{figure*}[!hbtp]
\centering
\subfigure{\resizebox{0.38\textwidth}{!}{\includegraphics[angle=0,scale=1.0]{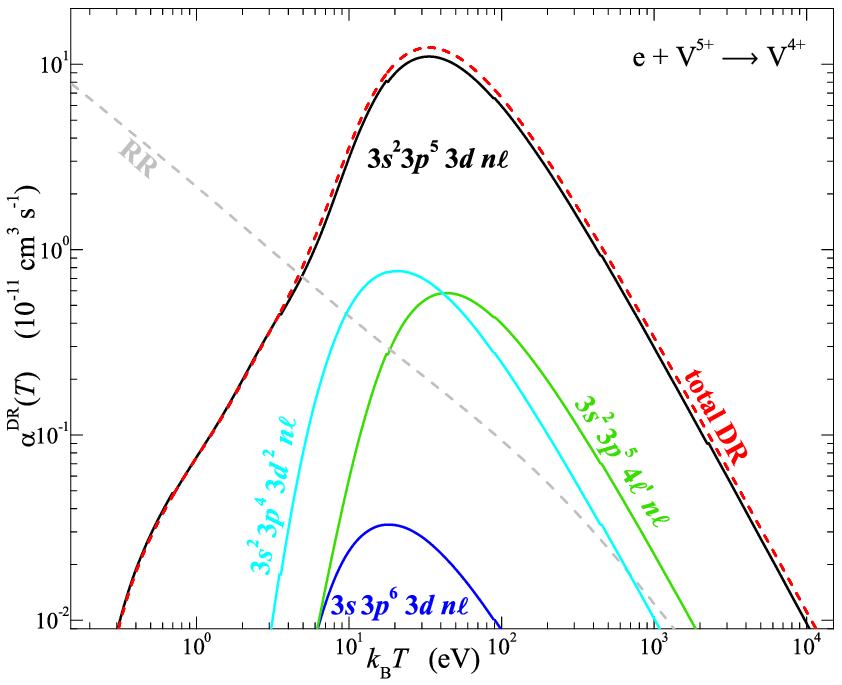}  }\label{Fig:V:ASMaxwl} }
\subfigure{\resizebox{0.38\textwidth}{!}{\includegraphics[angle=0,scale=1.0]{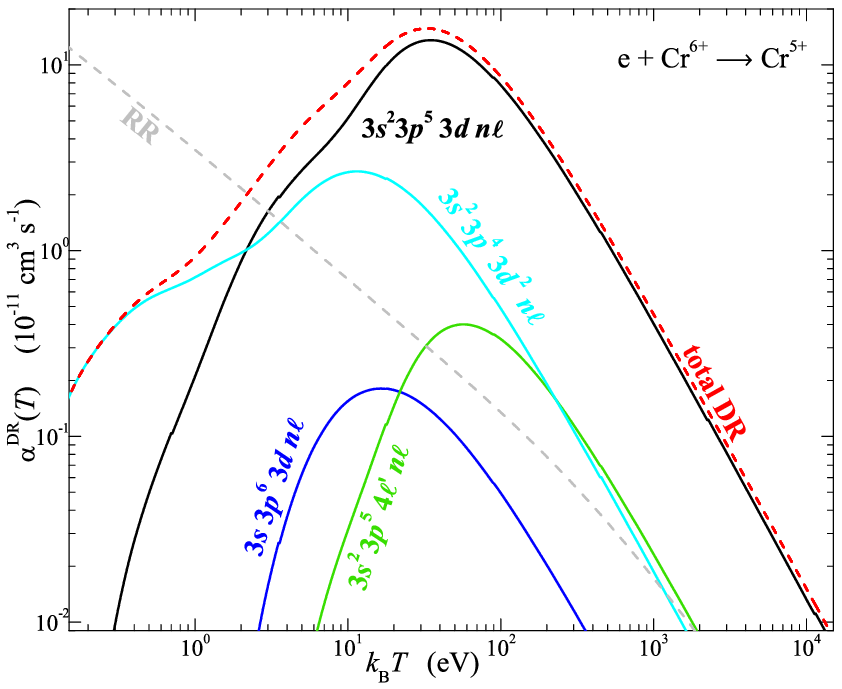} }\label{Fig:Cr:ASMaxwl}}
\subfigure{\resizebox{0.38\textwidth}{!}{\includegraphics[angle=0,scale=1.0]{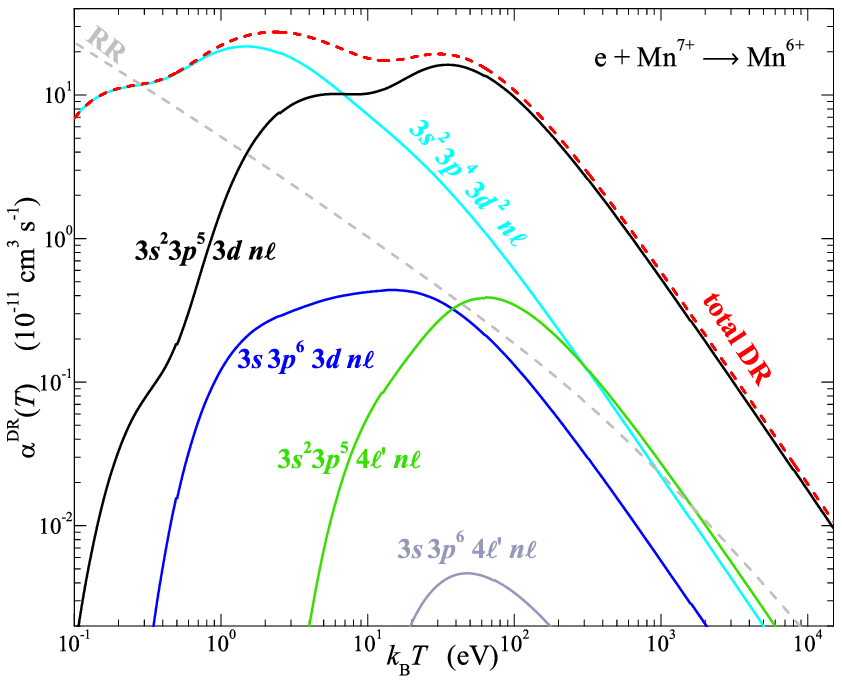} }\label{Fig:Mn:ASMaxwl}}
\subfigure{\resizebox{0.38\textwidth}{!}{\includegraphics[angle=0,scale=1.0]{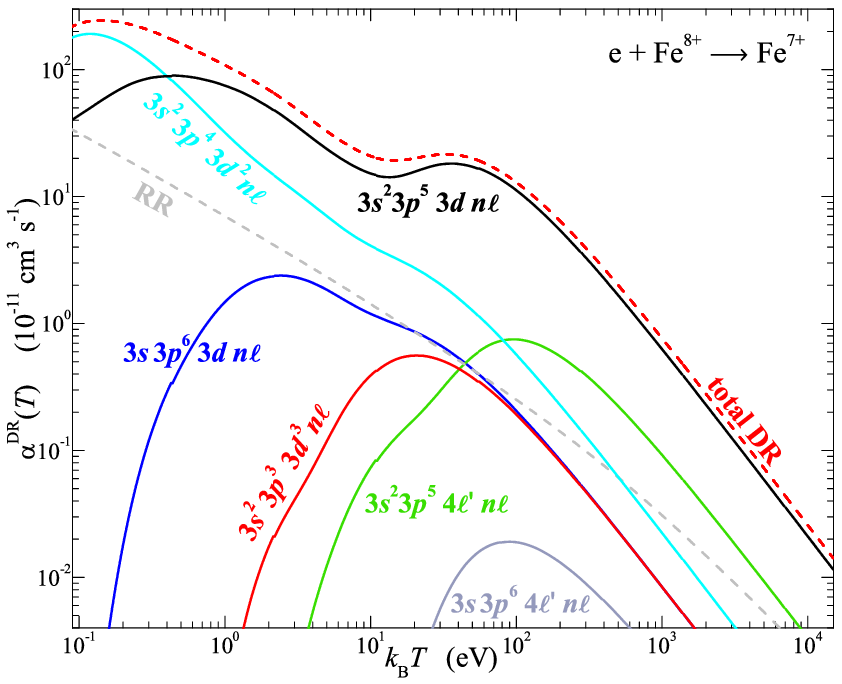} }\label{Fig:Fe:ASMaxwl}}
\subfigure{\resizebox{0.38\textwidth}{!}{\includegraphics[angle=0,scale=1.0]{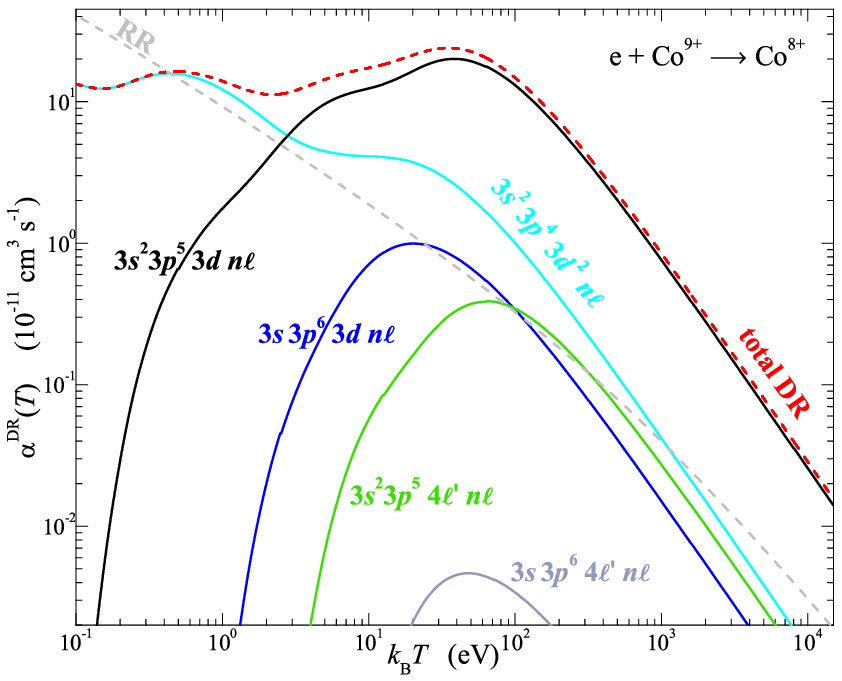} }\label{Fig:Co:ASMaxwl}}
\subfigure{\resizebox{0.38\textwidth}{!}{\includegraphics[angle=0,scale=1.0]{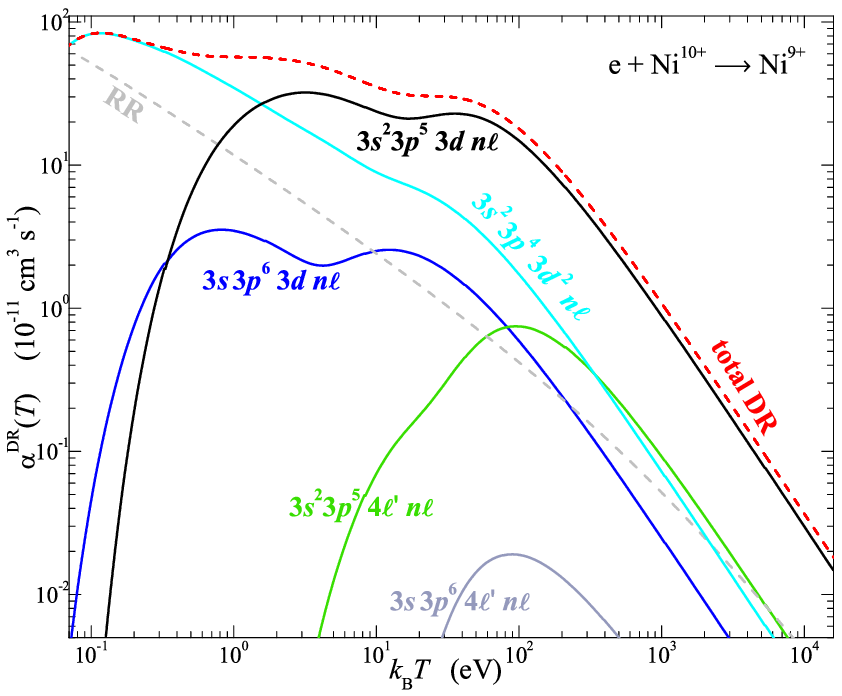}}\label{Fig:Ni:ASMaxwl}}
\subfigure{\resizebox{0.38\textwidth}{!}{\includegraphics[angle=0,scale=1.0]{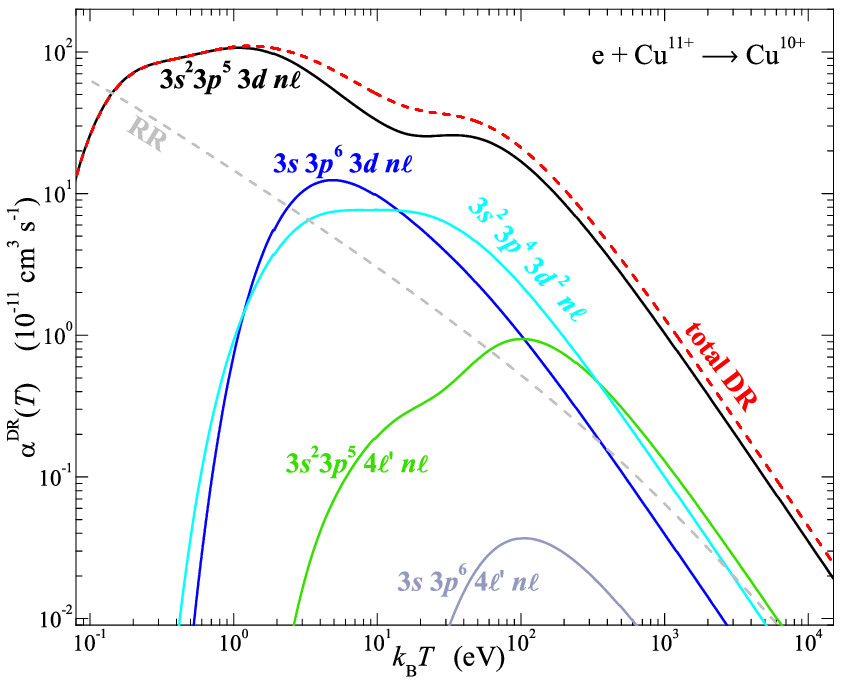}}\label{Fig:Cu:ASMaxwl}}
\subfigure{\resizebox{0.38\textwidth}{!}{\includegraphics[angle=0,scale=1.0]{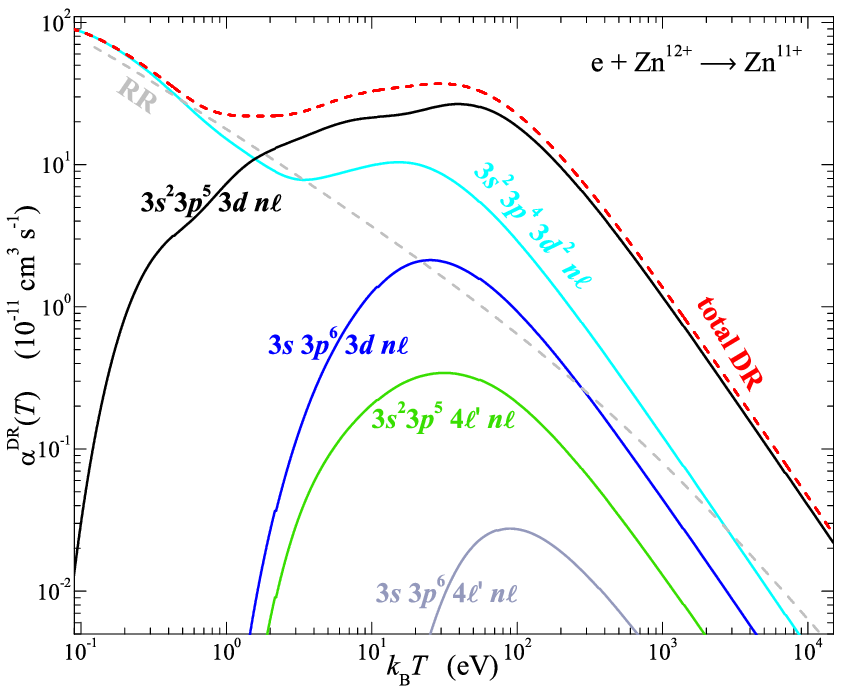}}\label{Fig:Zn:ASMaxwl}}
\caption[]{\label{Fig:V:Zn:ASMaxwl}
Present Maxwellian-averaged DR and RR rate coefficients for \ion{V}{vi} - \ion{Zn}{xiii} ions.
The red and gray dashed curves are the total DR and RR rate coefficients, respectfully.  The intra-shell $\Delta n_{{\rm c}} = 0$
contributions are due to $3p\to 3d$ (solid black curve), $3s\to 3d$ (solid blue curve), and $3p^2\to 3d^2$ (solid cyan curve) core excitations;
the inter-shell $\Delta n_{{\rm c}} = 1$ contributions are due to $3p\to 4\ell'$ (solid green curve)
and $3s\to 4\ell'$ (solid gray curve) promotions.
}
\end{figure*}

To treat the DR processes occurring in Eq.~(\ref{Eq:process}), all possible continuum and resonance wavefunctions
are constructed by coupling an appropriate distorted-wave free $\epsilon\ell$ or bound $n\ell$ orbital to each target
configuration wavefunction, as obtained in the previous section.
The energies, radiative rates, and Auger rates of each resonance are then computed and used in
Eqs.~(\ref{Eq:DR-IPIR})~and~(\ref{Eq:MaxwRate}) to produce Maxwellian rate coefficients that are shown in
Figs.~\ref{Fig:K:Ca:ASMaxwl}, \ref{Fig:Sc:Ti:ASMaxwl}, and \ref{Fig:V:Zn:ASMaxwl}.
In all the figures, the bars indicating the collisionally ionized zone were obtained assuming the
conditions of coronal equilibrium with electron-collisional plasma temperatures
for which the fractional abundance of the target ion in question surpasses 10\% of its peak value \citep{Bryans:2008}.
Similar approximate temperature indicators for gas in photoionization equilibrium have been computed using Cloudy
(v08.00; \cite{CLOUDY:90}). Although these results did not include the DR rate coefficient data reported here,
the indicated temperature range should still be reasonably accurate.

\begin{figure*}[!htbp]
\centering
\subfigure{\resizebox{0.48\textwidth}{!}{\includegraphics[angle=0,scale=1.0]{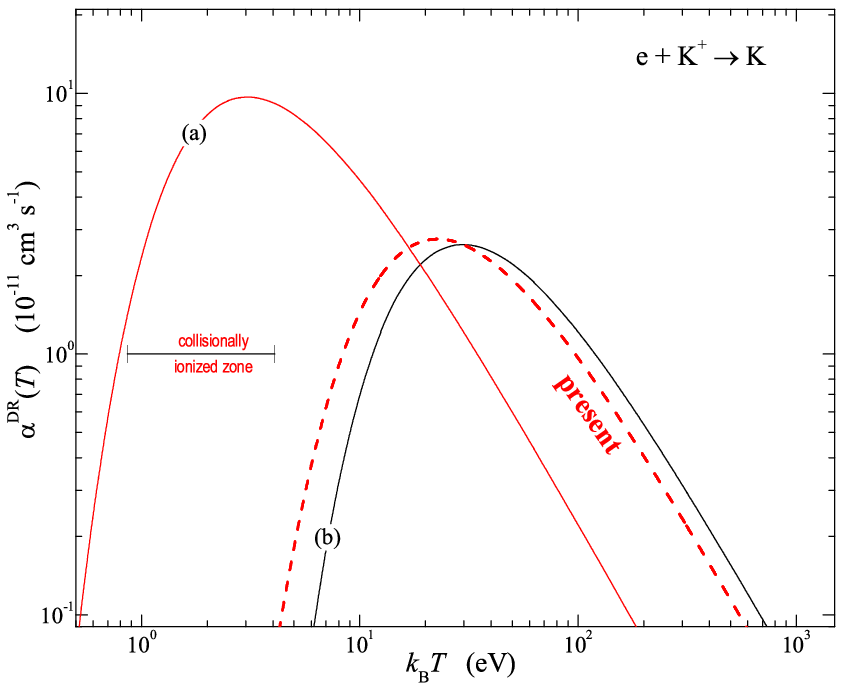}} \label{Fig:K:OldMaxwl}}
\subfigure{\resizebox{0.48\textwidth}{!}{\includegraphics[angle=0,scale=1.0]{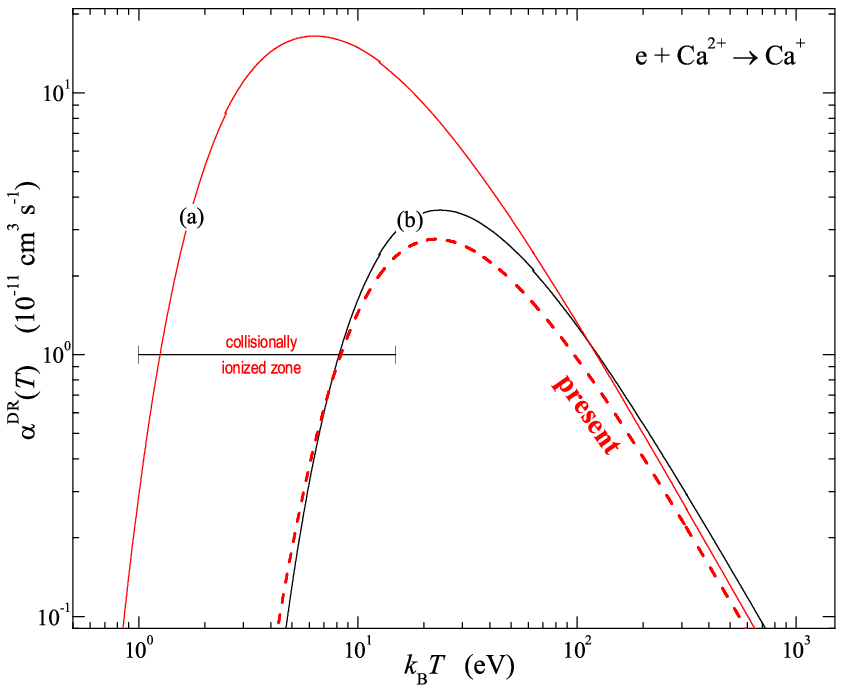}}\label{Fig:Ca:OldMaxwl}}
\caption[]{\label{Fig:K:Ca:OldMaxwl}
Comparison of existing total DR ground-level rate coefficients for \ion{K}{ii} (left) and \ion{Ca}{iii} (right):
(a) red solid curve, empirical results of \cite{Mewe:1980};
(b) black solid curve, recommended data of \cite{Mazzotta:1998};
    red dashed curve, present MCBP results.
}
\end{figure*}

For \ion{K}{ii} and \ion{Ca}{iii}, the only other data available in Fig.~\ref{Fig:K:Ca:OldMaxwl} are the results of
\cite{Mewe:1980} and \cite{Mazzotta:1998}. The rate coefficients of \cite{Mazzotta:1998} are inferred by using the general
formula of \cite{Burgess:1965} that is first scaled down to match the empirical DR rate of \cite{Hahn:1989} for \ion{Fe}{ix}
and then this same scaling is applied to other ions in the isoelectronic sequence.
Surprisingly, this procedure gives results that are in fair agreement with our present calculations, but, as is seen for
higher ionization stages, this method fails to account correctly for low-temperature DR. The results of \cite{Mewe:1980},
on the other hand, were obtained using a parameterized empirical formula based upon the renormalized results of
\cite{Jacobs:1977} and \cite{Ansari:1970} for $3p\to 3d$, $3p\to 4s$, and $3p\to 4d$ core-excitations of \ion{Fe}{ix};
that procedure gives erroneous results for the position and peak of the rate coefficient.

\begin{figure*}[!hbtp]
\centering
\subfigure{\resizebox{0.48\textwidth}{!}{\includegraphics[angle=0,scale=1.0]{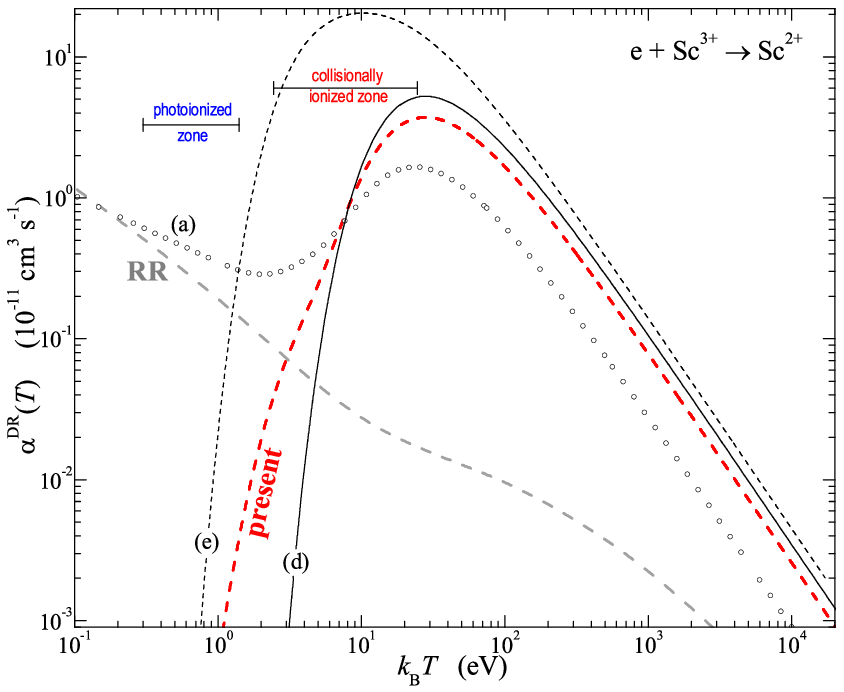} }\label{Fig:Sc:OldMaxwl}}
\subfigure{\resizebox{0.48\textwidth}{!}{\includegraphics[angle=0,scale=1.0]{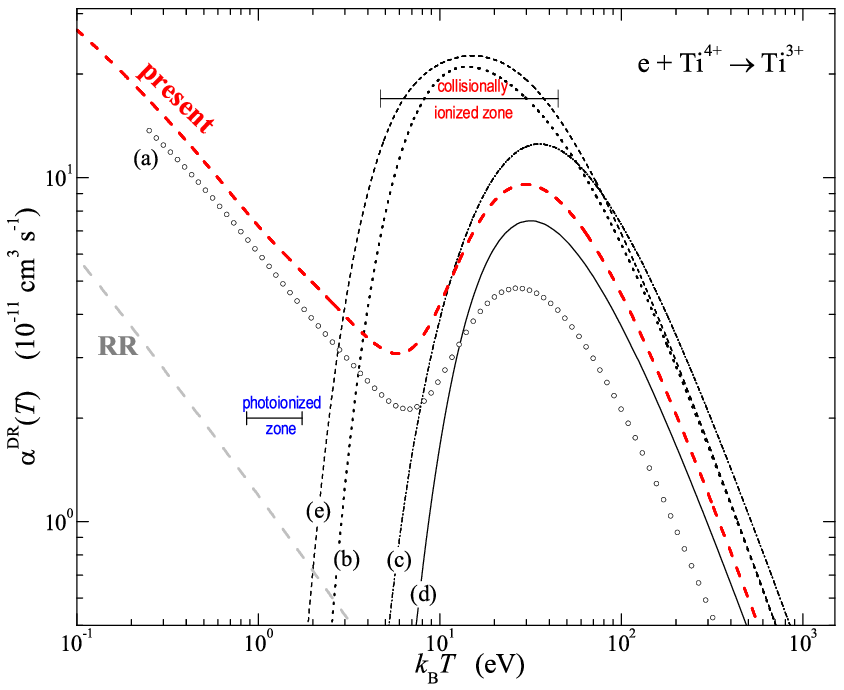} }\label{Fig:Ti:OldMaxwl}}
\caption[]{\label{Fig:Sc:Ti:OldMaxwl}
Comparison of existing total DR ground-level rate coefficients for \ion{Sc}{iv} (left) and \ion{Ti}{v} (right):
(a) gray open circles, TSR experiments by \cite{Schippers:1998,Schippers:2002};
(b) black doted curve, compilation by \cite{Landini:1991};
(c) black dash-dotted curve, empirical formula of \cite{Hahn:1991};
(d) black solid curve, recommended value by \cite{Mazzotta:1998};
(e) black dashed curve, empirical results of \cite{Mewe:1980}.
The present RR results are also shown as the long-dashed curve.
}
\end{figure*}

The results shown in Fig.~\ref{Fig:Sc:Ti:OldMaxwl} have been analyzed more fully in our earlier studies of \ion{Sc}{iv}
\citep{Nikolic:2010} and \ion{Ti}{v} \citep{Nikolic:2009}, but the following points should be made. First, there is
experimental data available for both of these ions \citep{Schippers:1998,Schippers:2002}, and these measurements were
useful for quantifying the positions of low-energy resonances. However, the experiments for these low-charged Ar-like
ions were subject to motional Stark effect that reionized the higher-$n$ recombined bound states and led to a reduction
in the DR cross section. Thus, the measured rate coefficient is unphysically too low at higher temperatures and is
inappropriate for use in plasma modeling.  Also, the experimental rate coefficient also contains the contribution due to RR,
which is why the \ion{Sc}{iv} experimental rate coefficient exceeds our DR result at lower temperatures and merges with
our RR result.

Another new feature seen in Fig.~\ref{Fig:Sc:Ti:OldMaxwl} that was absent in Fig.~\ref{Fig:K:Ca:OldMaxwl} is that, as the
ionization stage is increased, the lowest-lying members of the Rydberg series given in Eq.~(\ref{Eq:process}), namely the
$3s^{2}3p^{5}3d^{2}$ and $3s^{2}3p^{5}3d4s$ resonances, approach zero continuum energy and give a large contribution to
the rate coefficient at the lower temperatures found in photoionized plasmas. This contribution is not included in the
results of \cite{Mazzotta:1998}; those data are determined from the Burgess formula (\cite{Burgess:1965}), which only
considers high-temperature DR. We note that the data of \cite{Mazzotta:1998} also differs significantly from the present
DR rate coefficient in the collisionally ionized zone. The other available data - the empirical results of \cite{Mewe:1980},
\cite{Landini:1991}, and \cite{Hahn:1991} - differ dramatically from the present MCBP results.

\begin{figure*}[!htbp]
\centering
\subfigure{\resizebox{0.48\textwidth}{!}{\includegraphics[angle=0,scale=1.0]{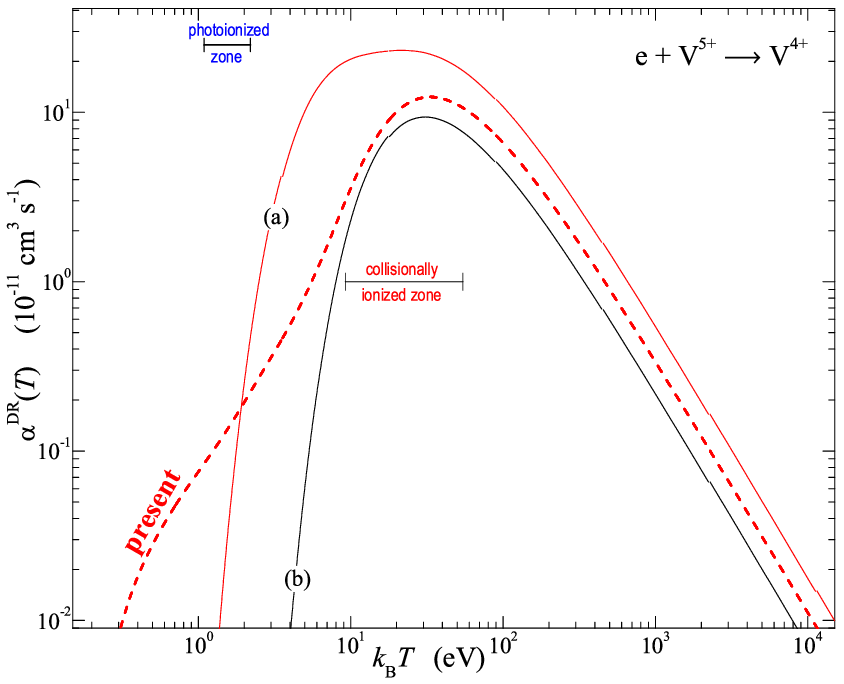}} \label{Fig:V:OldMaxwl}}
\subfigure{\resizebox{0.48\textwidth}{!}{\includegraphics[angle=0,scale=1.0]{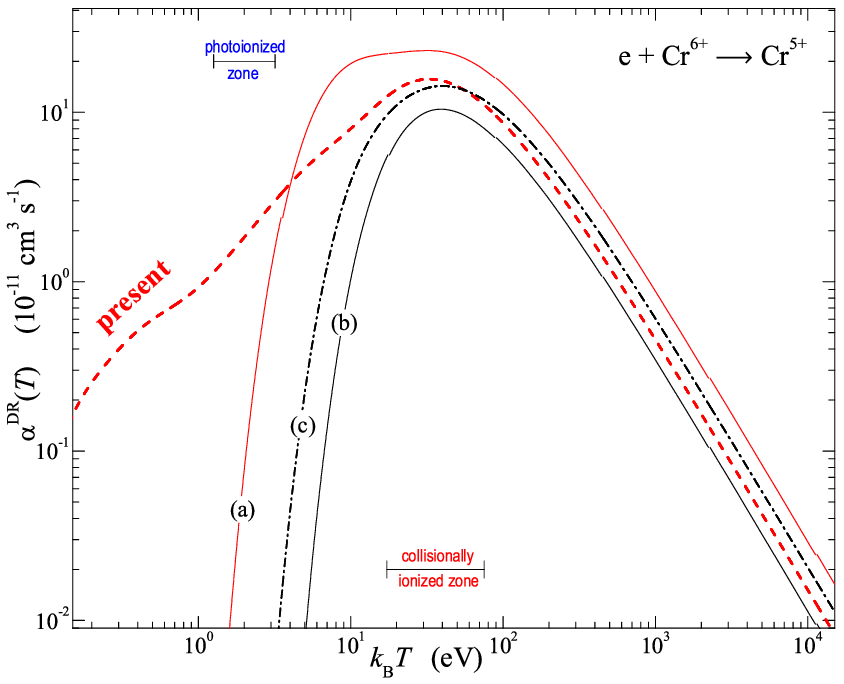}}\label{Fig:Cr:OldMaxwl}}
\caption[]{\label{Fig:V:Cr:OldMaxwl}
Comparison of existing total DR ground-level rate coefficients for \ion{V}{vi} (left) and \ion{Cr}{vii} (right):
(a) red solid curve, empirical results of \cite{Mewe:1980};
(b) black solid curve, recommended data of \cite{Mazzotta:1998};
(c) black dash-dotted curve, empirical formula of \cite{Hahn:1991};
    red dashed curve, present MCBP results.
}
\end{figure*}

\begin{figure*}[!hbtp]
\centering
\subfigure{\resizebox{0.48\textwidth}{!}{\includegraphics[angle=0,scale=1.0]{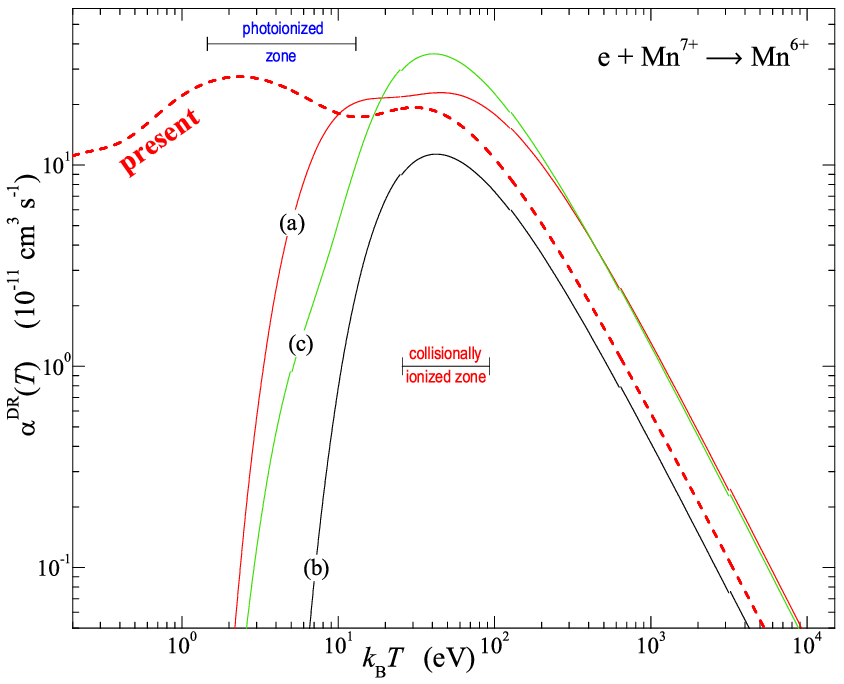}}\label{Fig:Mn:OldMaxwl}}
\subfigure{\resizebox{0.48\textwidth}{!}{\includegraphics[angle=0,scale=1.0]{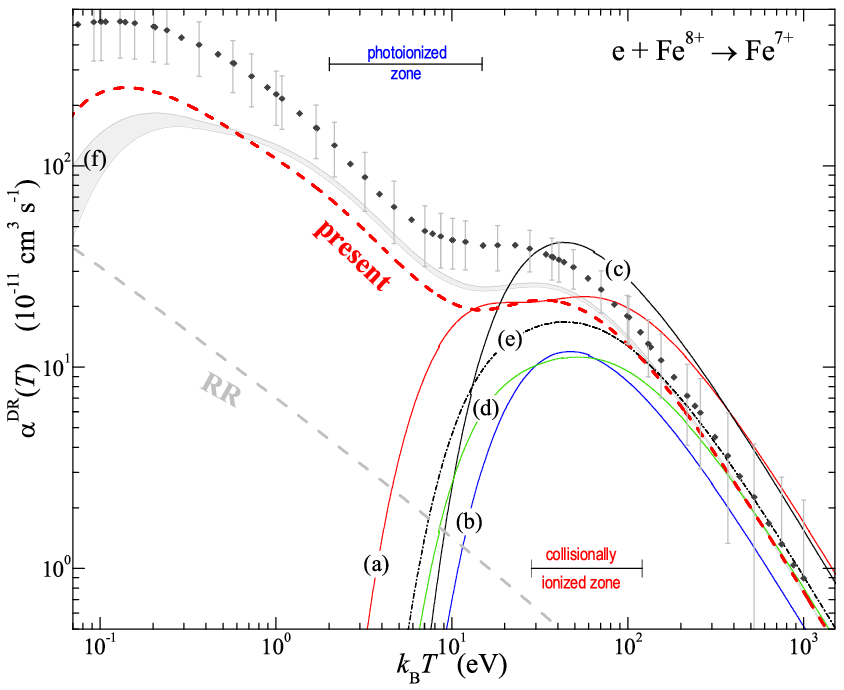}}\label{Fig:Fe:OldMaxwl}}
\caption[]{\label{Fig:Mn:Fe:OldMaxwl}
Comparison of existing total DR ground-level rate coefficients. On the left  is for \ion{Mn}{viii}:
(a) red solid curve, empirical results of \cite{Mewe:1980};
(b) black solid curve, recommended data of \cite{Mazzotta:1998};
(c) green solid curve, LS results of \cite{Jacobs:1983};
    red dashed curve, present MCBP results.
On the right  is \ion{Fe}{ix}:
(a) red solid curve, empirical results of \cite{Mewe:1980};
(b) blue solid curve, compilation of \cite{Arnaud:1992};
(c) black solid curve, result of \cite{Jacobs:1977};
(d) green solid curve, empirical data of \cite{Kato:1999};
(e) black dash-dotted curve, empirical formula of \cite{Hahn:1991};
(f) "4CF" and "5CF" MCBP results of \cite{Badnell:Fe3pq:2006};
    red dashed curve, present "18CF" MCBP results.
The present RR results are also shown as the long-dashed curve.
}
\end{figure*}

\begin{figure*}[!hbtp]
\centering
\subfigure{\resizebox{0.48\textwidth}{!}{\includegraphics[angle=0,scale=1.0]{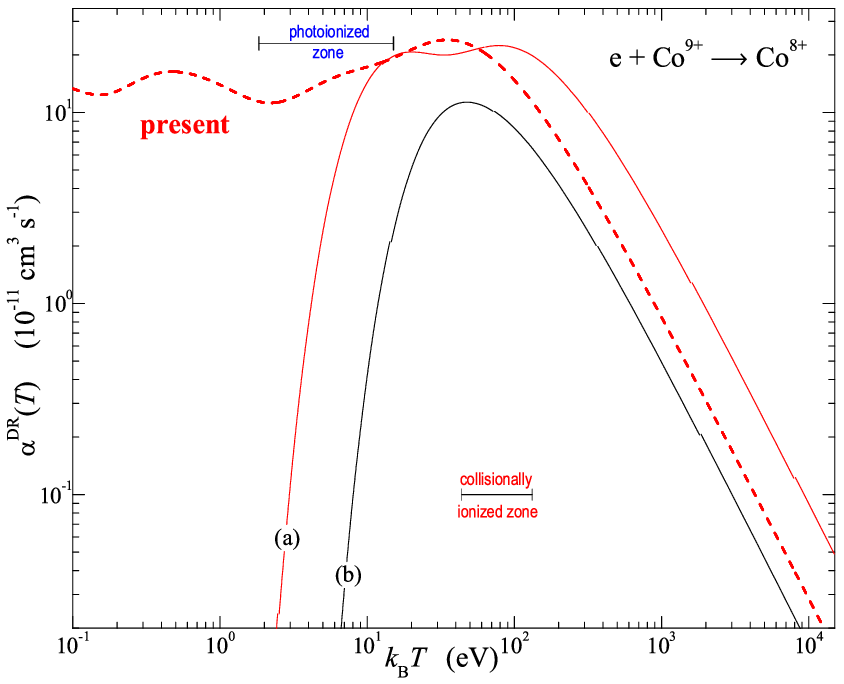}} \label{Fig:Co:OldMaxwl}}
\subfigure{\resizebox{0.48\textwidth}{!}{\includegraphics[angle=0,scale=1.0]{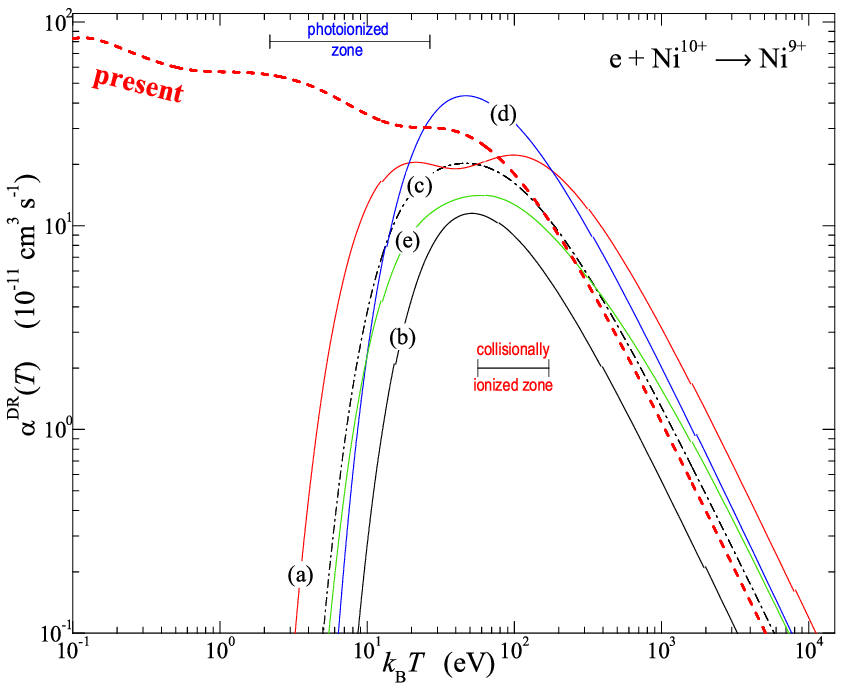}}\label{Fig:Ni:OldMaxwl}}
\caption[]{\label{Fig:Co:Ni:OldMaxwl}
Comparison of existing total DR ground-level rate coefficients. On the left is for \ion{Co}{x}:
(a) red solid curve, empirical results of \cite{Mewe:1980};
(b) black solid curve, recommended data of \cite{Mazzotta:1998};
    red dashed curve, present MCBP results.
On the right is for \ion{Ni}{xi} ion:
(a) red solid curve, empirical results of \cite{Mewe:1980};
(b) black solid curve, recommended data of \cite{Mazzotta:1998};
(c) black dash-doted curve, empirical formula of \cite{Hahn:1991};
(d) blue solid curve, compilation of \cite{Arnaud:1985};
(e) green solid curve, empirical data of \cite{Kato:1999};
    red dashed curve, the present MCBP results.
}
\end{figure*}

\begin{figure*}[!hbtp]
\centering
\subfigure{\resizebox{0.48\textwidth}{!}{\includegraphics[angle=0,scale=1.0]{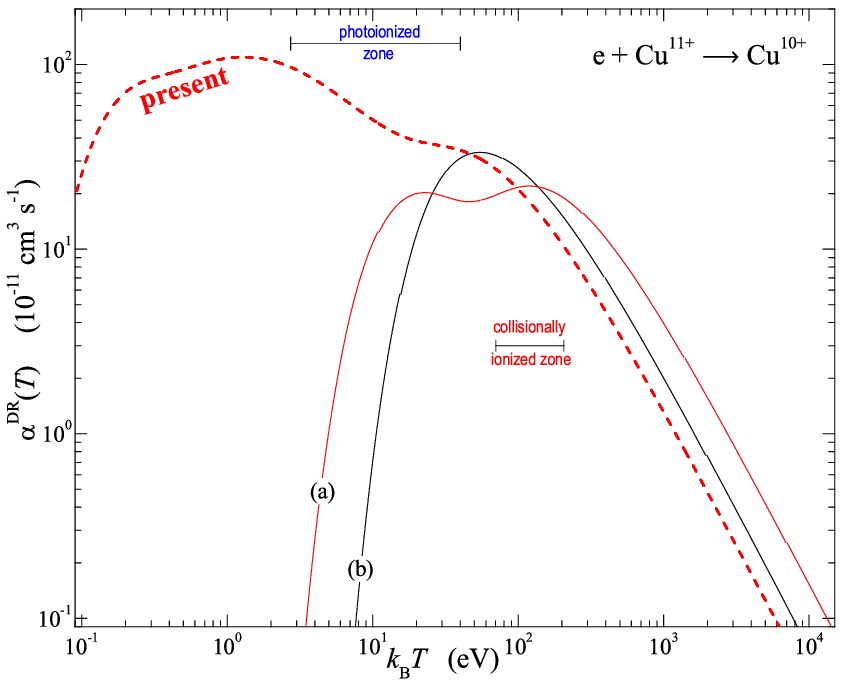}} \label{Fig:Cu:OldMaxwl}}
\subfigure{\resizebox{0.48\textwidth}{!}{\includegraphics[angle=0,scale=1.0]{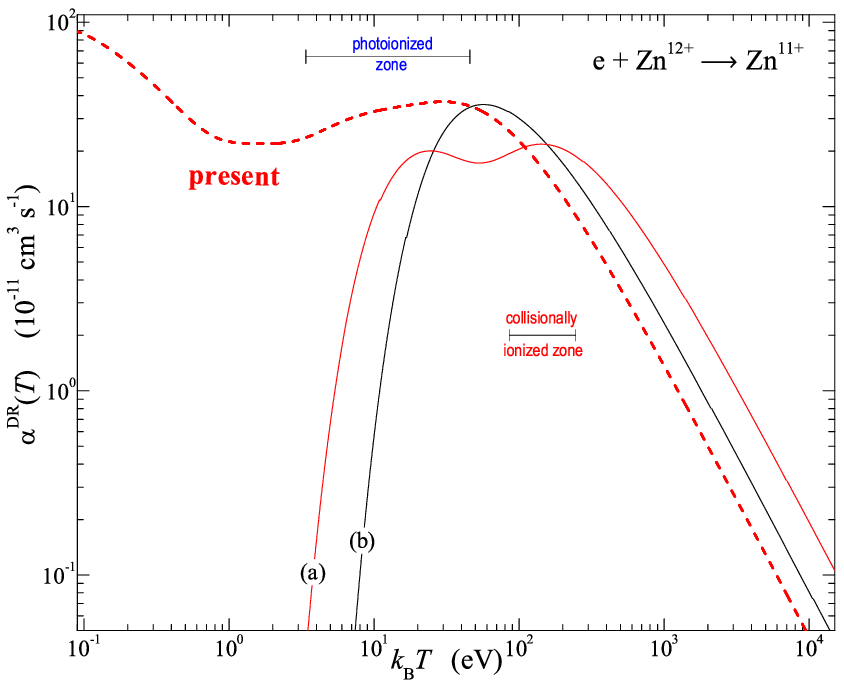}} \label{Fig:Zn:OldMaxwl}}
\caption[]{\label{Fig:Cu:Zn:OldMaxwl}
Comparison of existing total DR ground-level rate coefficients for \ion{Cu}{xii} (left) and \ion{Zn}{xiii} (right):
(a) red solid curve, empirical results of \cite{Mewe:1980};
(b) black solid curve, recommended data of \cite{Mazzitelli:2002};
    red dashed curve, present MCBP results.
}
\end{figure*}

For the higher ionization stages of \ion{V}{vi}, \ion{Cr}{vii}, and \ion{Mn}{viii} shown in
Figs.~\ref{Fig:V:Cr:OldMaxwl}~and~\ref{Fig:Mn:Fe:OldMaxwl}, the present rate coefficient again has a significant
low-temperature DR contribution that is absent in all previously available data. For \ion{Fe}{ix}, there exist several
other results, but most of these are again empirical results that fail to account for low-temperature DR contributions
in the photoionized plasma region and also differ significantly from our present DR results in the collisionally-ionized
plasma region. We also show the earlier MCBP results of \cite{Badnell:Fe3pq:2006}. Those calculations were quite similar
to the present ones, except that a smaller configuration basis was used and thus a slightly larger rate coefficient was
obtained. We have established that by augmenting those earlier calculations with the inclusion of additional
correlation configurations, as we include here, the two results are brought into agreement. Both MCBP results are somewhat
lower than the experimental rate coefficient, as has been discussed more fully by \cite{Schmidt:2008}. We note that the
experiment for \ion{Fe}{ix} was not influenced by external fields, unlike the experiments for \ion{Sc}{iv} and \ion{Ti}{v}.
Our DR rate coefficients for \ion{Co}{x} through \ion{Zn}{xiii} are shown in
Figs.~\ref{Fig:Co:Ni:OldMaxwl}~and~\ref{Fig:Cu:Zn:OldMaxwl}, where it is seen again that all previous data do not account
for the low-temperature contributions that dominate in the photoionized plasma region.

In Figs.~\ref{Fig:K:Ca:ASMaxwl}, \ref{Fig:Sc:Ti:ASMaxwl}, and \ref{Fig:V:Zn:ASMaxwl}, we show the DR contributions from
each resonance series. As anticipated due to its large core oscillator strength, the $3s^{2}3p^{5}3dn\ell$ resonances are
dominant. The lowest $3s^{2}3p^{5}3d^{2}$ resonances are first seen to be indistinct from the rest of the series for
\ion{K}{ii} and \ion{Ca}{iii}, then begin to show a separate feature at low temperatures for \ion{Sc}{iv}, and completely
dominate at low temperatures for \ion{Ti}{v}. As the nuclear charge is increased from \ion{V}{vi} through \ion{Zn}{xiii},
additional low-temperature features are seen to oscillate as the $n=4$ and $n=5$ resonances move from above to below threshold.
Also seen as the ionization stage increases is the appearance of the $3s^{2}3p^{4}3d^{2}n\ell$ ($n=3,4$) resonances near
threshold that dominate the low-temperature rate coefficient.

Lastly, we list the DR fitting coefficients for each ion, as described by Eq.~(\ref{Eq:Fit:DR}), in Table~\ref{Tab:DR:Maxwl}.
The RR fitting coefficients of Eq.~(\ref{Eq:Fit:RR}) are listed in Table~\ref{Tab:RR:Maxwl}.

\section{Summary}
\label{Sec:Summary}

It is clear that all previously-available DR data for Ar-like ions are inadequate at lower temperatures since the
contributions from low-lying resonances are not accounted for. Thus, those earlier data are inappropriate for use
in modeling photoionized plasmas. And even at higher temperatures relevant to collisionally ionized plasmas, the
earlier data were obtained using rather crude parameterization formulas that are not as reliable as our computed
MCBP rate coefficients. The results reported here serve as what we assess to be the most accurate and complete set
of Ar-like DR data for use in spectral diagnostic studies of laboratory and astrophysical plasmas.

\begin{acknowledgements}
This work was funded in part by NASA APRA, NASA SHP SR\&T, and PPARC grants.
\end{acknowledgements}

{\tiny

}

\clearpage

\clearpage

\clearpage

\clearpage

\clearpage


\clearpage
\begin{table}[!htbp]
\begin{minipage}[t]{\textwidth}
\caption{\label{Tab:lambdas}
Radial scaling parameters, $\lambda_{nl}$, for the $n=3$ valence orbitals optimized in the Slater-type-orbital model potential.
Details about the full optimization procedure are reported earlier by \cite{Nikolic:2009}.
}
\resizebox{\textwidth}{!}{
\begin{tabular}{c l l l l l l l l l l l l}    
\hline
\hline
\multicolumn{11}{c}{\vspace*{-2mm}} \\
&${\rm K}^{+}$&${\rm Ca}^{2+}$&${\rm Sc}^{3+,{\rm a}}$&${\rm Ti}^{4+,{\rm b}}$&${\rm V}^{5+}$&${\rm Cr}^{6+}$&${\rm Mn}^{7+}$&${\rm Fe}^{8+}$&${\rm Co}^{9+}$&${\rm Ni}^{10+}$&${\rm Cu}^{11+}$&${\rm Zn}^{12+}$ \\
\hline
\multicolumn{11}{c}{\vspace*{-2mm}} \\
$\lambda_{3s}$ & $1.09548$ & $1.09548$ & $1.08366$ & $1.05653$ &$1.07085$ & $1.07085$ & $1.07085$ & $1.05990$ & $1.0000$ & $1.0000$ & $1.0000$ & $1.0000$ \\
$\lambda_{3p}$ & $1.11185$ & $1.10545$ & $0.97900$ & $0.98950$ &$1.14255$ & $1.11729$ & $1.09171$ & $1.07936$ & $1.0693$ & $1.0608$ & $1.0453$ & $1.0318$ \\
$\lambda_{3d}$ & $1.13830$ & $1.16918$ & $1.27700$ & $1.17400$ &$0.89575$ & $0.87575$ & $0.87575$ & $0.84905$ & $0.8391$ & $0.8191$ & $0.8191$ & $0.8191$ \\
\hline
\end{tabular}
} 
    adopted from: $^{{\rm a} }$ \cite{Nikolic:2010}; $^{{\rm b} }$ \cite{Nikolic:2009}.
\end{minipage}
\end{table}

\begin{table}[!htbp]
\begin{minipage}[t]{\textwidth}
\caption{\label{Tab:tar-rad}
Dominant radiative transition data from the ground state of the
recombining ion $3p^6\;({}^1S_{0})\to 3p^53d\;({}^1P_{1}^{{\rm o}})$, where $v_{(u)}^{[\pm p]}$ denotes $v(u)\times 10^{\pm p}$.
In the case of ${\rm Sc}^{3+}$ and ${\rm Ti}^{4+}$ ions, refer to \cite{Nikolic:2010} and \cite{Nikolic:2009}, respectively.
}
\resizebox{\textwidth}{!}{
\begin{tabular}{c l l l l c l l l l}   
\hline
\hline
\multicolumn{10}{c}{\vspace*{-2mm}} \\
$\begin{array}{c} {\rm Ion} \\  \end{array}$ &
$\begin{array}{c} A^{r} \\ ({\rm ns}^{-1}) \end{array}$ &
$\begin{array}{c} S \\ ({\rm a.u.}) \end{array}$ &
$\begin{array}{c} f \\  \hfill \end{array}$ &
$\begin{array}{c} {\rm Ref.} \\ \hfill \end{array}$ &
$\begin{array}{c} {\rm Ion} \\  \end{array}$ &
$\begin{array}{c} A^{r} \\ ({\rm ns}^{-1}) \end{array}$ &
$\begin{array}{c} S \\ ({\rm a.u.}) \end{array}$ &
$\begin{array}{c} f \\  \hfill \end{array}$ &
$\begin{array}{c} {\rm Ref.} \\ \hfill \end{array}$ \\
\hline
\multicolumn{10}{c}{\vspace*{-2mm}} \\
\multirow{2}{*}{${\rm K}^{+}$}&$0.2785^{[+2]}$&$5.004$&$3.070$&Present$^{{\rm a}}$&\multirow{4}{*}{${\rm Ca}^{2+}$}&$0.836^{[+2]}$&$5.678$&$4.818$& Present$^{{\rm a}}$\\
               & $0.2224^{[+2]}$ & $4.147$ & $2.513$ & MCHF$^{{\rm c}}    $ &   &$0.83^{[+2]}_{(15)}$&$5.64$&$4.75_{(88)}$&NIST$^{{\rm b},\,\S}$\\
\cline{2-5}
\multirow{3}{*}{${\rm V}^{5+}$}&$1.4610^{[+2]}$&$2.448$&$3.313$&Present$^{{\rm a}}$&        &$0.638^{[+2]}$&$4.614$&$3.834$&MCHF$^{{\rm c}}    $\\
               &$1.46  ^{[+2]} $ & $2.45 $ & $3.31 $ & NIST$^{{\rm b}\dag}$ &               &$0.682^{[+2]}$&$4.210$&$3.688$&CIV3$^{{\rm j}}    $\\
\cline{7-10}
               &$1.8447^{[+2]}$&$2.968$&$4.070$&MCHF$^{{\rm c}}$&\multirow{3}{*}{${\rm Cr}^{6+}$}&$1.65^{[+2]}$&$2.04$&$3.05$&Present$^{{\rm a}}$\\
\cline{2-5}
\multirow{3}{*}{${\rm Mn}^{7+}$}&$1.858^{[+2]}$&$1.754$&$2.874$&Present$^{{\rm a}}$&        &$1.67^{[+2]}$&$2.06 $ &$ 3.09$&NIST$^{{\rm b}\ddag}$\\
               &$1.85 ^{[+2]}  $ &$1.75 $  &$2.86 $  &NIST$^{{\rm b}\ddag}$ &               &$2.05^{[+2]}$&$2.43 $ &$ 3.69$&MCHF$^{{\rm c}}     $\\
\cline{7-10}
         &$1.963^{[+2]}$ &$2.033$ &$3.229$ &MCHF$^{{\rm c}}$ &\multirow{8}{*}{${\rm Fe}^{8+}$}&$2.009^{[+2]}$&$1.489$&$2.645$&Present$^{{\rm a}}$\\
\cline{2-5}
\multirow{4}{*}{${\rm Co}^{9+}$}&$2.190^{[+2]}$&$1.300$&$2.486$&Present$^{{\rm a}}$&        &$2.01 ^{[+2]}$&$1.49 $&$2.65 $&NIST $^{{\rm b}\dag}$\\
               &$2.200^{[+2]}  $ &$1.300 $ &$2.500 $ &NIST$^{{\rm b}\ddag}$ &               &$2.097^{[+2]}$&$1.724$&$2.958$&MCHF $^{{\rm c}}    $\\
               &$2.224^{[+2]}  $ &$1.482 $ &$2.726 $ &MCHF$^{{\rm c}}     $ &               &$2.438^{[+2]}$&$1.678$&$3.054$&FAC4 $^{{\rm d}}    $\\
               &$1.448^{[+2]}  $ &$1.409 $ &$2.628 $ &TDCHF$^{{\rm f}}    $ &               &$2.266^{[+2]}$&$1.680$&$2.983$&CIV3 $^{{\rm e}}    $\\
\cline{2-5}
\multirow{6}{*}{${\rm Ni}^{10+}$}&$2.339^{[+2]}$&$1.132$&$2.317$&Present$^{{\rm a}}$&       &$1.356^{[+2]}$&$1.073$&$2.843$&TDCHF$^{{\rm f}}    $\\
                &$2.340^{[+2]} $ &$1.130 $ &$2.320 $ &NIST$^{{\rm b}\dag} $ &               &$2.889^{[+2]}$&       &       &SELT $^{{\rm g}}    $\\
                &$2.345^{[+2]} $ &$1.288 $ &$2.528 $ &MCHF$^{{\rm c}}     $ &               &$2.404^{[+2]}$&$1.637$&$2.99 $&IRON $^{{\rm h}}    $\\
\cline{7-10}
            &$1.537^{[+2]}$&$0.808$&$2.444$&TDCHF$^{{\rm f}}$&\multirow{4}{*}{${\rm Cu}^{11+}$}&$2.530^{[+2]}$&$1.01$&$2.204$&Present$^{{\rm a}}$\\
                &$2.786^{[+2]} $ &$1.274 $ &$2.657 $ &FAC3$^{{\rm h}}     $ &  &$2.501^{[+2]}_{(30)}$&$1.07_{(11)}$&$2.29_{(7)}$&NIST$^{{\rm b},\,\S}$\\
                &$2.841^{[+2]} $ &$1.204 $ &$2.575 $ &CIV3$^{{\rm i}}     $ &                &$2.462^{[+2]}$&$1.13$&$2.355  $&MCHF$^{{\rm c}}$\\
\cline{2-5}
\multirow{4}{*}{${\rm Zn}^{12+}$}&$2.720^{[+2]}$&$0.906$&$2.101$&Present$^{{\rm a}}$&        &$1.624^{[+2]}$&$1.08$&$2.285  $&TDCHF$^{{\rm f}}$\\
    &$2.653^{[+2]}_{(33)}$&$0.99_{(11)}$&$2.21_{(7)}$&NIST$^{{\rm b},\,\S}$&                &              &      &            &                   \\
                &$2.576^{[+2]} $ &$1.001 $ &$2.204 $ &MCHF$^{{\rm c}}     $&                &              &      &            &                   \\
                &$1.708^{[+2]} $ &$0.958 $ &$2.145 $ &TDCHF$^{{\rm f}}    $&                &              &      &            &                   \\
\hline
\hline
\end{tabular}
} 
    \begin{list}{}{}
    \item[$^{{\rm  } }$] assigned uncertainties: $^{\dag)}\,< 18\%$, $^{\ddag)}\,< 25\%$
    \item[$^{{\rm a} }$] present work: 2894-level MCBP results\
    \item[$^{{\rm b} }$] critically compiled experimental data of \cite{Shirai:2000}
    \item[$^{{\rm c} }$] MCHF results of \cite{Irimia:2003}
    \item[$^{{\rm d} }$] 6284-level FAC results of \cite{Aggarwal:2006}
    \item[$^{{\rm e} }$] CIV3 results of \cite{Verma:2006}
    \item[$^{{\rm f} }$] single-config. TDCHF results\ of \cite{Ghosh:1997}
    \item[$^{{\rm g} }$] semiempirical results of \cite{Loginov:2002}
    \item[$^{{\rm h} }$] 6164-level FAC results of \cite{Aggarwal:2008}
    \item[$^{{\rm i} }$] restricted CIV3 results of \cite{Verma:2007}
    \item[$^{{\rm j} }$] 18-config. CIV3 LS results of \cite{Baluja:1986}
    \item[$^{\S      }$] extrapolated along the isoelectronic sequence
    \end{list}
\end{minipage}
\end{table}

\begin{table*}[!hbtp]
\begin{minipage}[t]{\textwidth}
\caption{\label{Tab:K:Ca:tar-str}
The lowest $\Delta n_{{\rm c}} = 0, 1$ core excitation thresholds (in Rydbergs)
for \ion{K}{ii} and \ion{Ca}{iii}. For brevity, we show only the core excitations
that are below the dominant $3p^6\;({}^1S_{0})\to 3p^53d\;({}^1P_{1}^{{\rm o}})$ threshold.
The exception is the $3p^54d\;({}^1P_{1}^{{\rm o}})$ threshold in \ion{K}{ii} whose
configuration is marked as questionable by \cite{Sugar:1985} and only recently
properly identified by \cite{Pettersen:2007}.
}
\resizebox{\textwidth}{!}{
\begin{tabular}{r l l l r r c l l r r r}  
\hline
\hline
\multicolumn{12}{c}{\vspace*{-2mm}} \\
   &         & \multicolumn{4}{c}{${\rm K}^{+}$} &   &   & \multicolumn{4}{c}{${\rm Ca}^{2+}$} \\
\multicolumn{12}{c}{\vspace*{-2mm}} \\
\cline{3-6}\cline{9-12}
\multicolumn{12}{c}{\vspace*{-2mm}} \\
K  & Config. & Level(mix) & Present$^{{\rm a}}$ & NIST$^{{\rm b}}$ & MCHF$^{{\rm c}}$ &
K  & Config. & Level(mix) & Present$^{{\rm a}}$ & NIST$^{{\rm d}}$ & MCHF$^{{\rm c}}$ \\
\hline
\multicolumn{12}{c}{\vspace*{-2mm}} \\
 1 & $3s^{2}3p^{6}  $ & ${}^{1}\hspace*{-.5mm}S_{0}\,(96.3\%)$                       & $0.000000$ & $0.000000$ & $0.000000$ &
 1 & $3s^{2}3p^{6}  $ & ${}^{1}\hspace*{-.5mm}S_{0}\,(99.0\%)$                       & $0.000000$ & $0.000000$ & $0.000000$ \\
\multicolumn{12}{c}{\vspace*{-2mm}} \\
 2 & $3s^{2}3p^{5}4s$ & ${}^{3}\hspace*{-.5mm}P_{2}^{{\rm o}}\,(93.2\%)$             & $1.480381$ & $1.480834$ & $1.400644$ &
 2 & $3s^{2}3p^{5}3d$ & ${}^{3}\hspace*{-.5mm}P_{0}^{{\rm o}}\,(95.7\%)$             & $1.879530$ & $1.853273$ & $1.760169$ \\
 3 & $3s^{2}3p^{5}4s$ & ${}^{3}\hspace*{-.5mm}P_{1}^{{\rm o}}\,(59.7\%)$             & $1.487826$ & $1.487481$ & $1.408491$ &
 3 & $3s^{2}3p^{5}3d$ & ${}^{3}\hspace*{-.5mm}P_{1}^{{\rm o}}\,(95.6\%)$             & $1.884279$ & $1.857636$ & $1.764514$ \\
 4 & $3s^{2}3p^{5}3d$ & ${}^{3}\hspace*{-.5mm}P_{0}^{{\rm o}}\,(70.9\%)$             & $1.490116$ & $1.489303$ & $1.431384$ &
 4 & $3s^{2}3p^{5}3d$ & ${}^{3}\hspace*{-.5mm}P_{2}^{{\rm o}}\,(95.2\%)$             & $1.893907$ & $1.866664$ & $1.773134$ \\
\multicolumn{12}{c}{\vspace*{-2mm}} \\
 5 & $3s^{2}3p^{5}3d$ & ${}^{3}\hspace*{-.5mm}P_{1}^{{\rm o}}\,(59.5\%)$             & $1.499706$ & $1.498961$ & $1.430119$ &
 5 & $3s^{2}3p^{5}3d$ & ${}^{3}\hspace*{-.5mm}F_{4}^{{\rm o}}\,(95.1\%)$             & $1.952880$ & $1.934711$ & $1.844004$ \\
 6 & $3s^{2}3p^{5}3d$ & ${}^{3}\hspace*{-.5mm}P_{2}^{{\rm o}}\,(81.1\%)$             & $1.505571$ & $1.502934$ & $1.436726$ &
 6 & $3s^{2}3p^{5}3d$ & ${}^{3}\hspace*{-.5mm}F_{3}^{{\rm o}}\,(93.9\%)$             & $1.963210$ & $1.944456$ & $1.853118$ \\
 7 & $3s^{2}3p^{5}4s$ & ${}^{3}\hspace*{-.5mm}P_{0}^{{\rm o}}\,(82.4\%)$             & $1.506735$ & $1.504916$ & $1.414847$ &
 7 & $3s^{2}3p^{5}3d$ & ${}^{3}\hspace*{-.5mm}F_{2}^{{\rm o}}\,(94.0\%)$             & $1.972601$ & $1.953156$ & $1.861216$ \\
\multicolumn{12}{c}{\vspace*{-2mm}} \\
 8 & $3s^{2}3p^{5}4s$ & ${}^{1}\hspace*{-.5mm}P_{1}^{{\rm o}}\,(74.8\%)$             & $1.517409$ & $1.516872$ & $1.437439$ &
 8 & $3s^{2}3p^{5}3d$ & ${}^{1}\hspace*{-.5mm}D_{2}^{{\rm o}}\,(58.9\%)$             & $2.086507$ & $2.057880$ & $1.972081$ \\
\multicolumn{12}{c}{\vspace*{-2mm}} \\
 9 & $3s^{2}3p^{5}3d$ & ${}^{3}\hspace*{-.5mm}F_{4}^{{\rm o}}\,(82.4\%)$             & $1.547896$ & $1.549602$ & $1.480702$ &
 9 & $3s^{2}3p^{5}3d$ & ${}^{3}\hspace*{-.5mm}D_{3}^{{\rm o}}\,(53.7\%)$             & $2.080602$ & $2.062503$ & $1.976286$ \\
10 & $3s^{2}3p^{5}3d$ & ${}^{3}\hspace*{-.5mm}F_{3}^{{\rm o}}\,(81.1\%)$             & $1.555741$ & $1.556728$ & $1.488382$ &
10 & $3s^{2}3p^{5}3d$ & ${}^{3}\hspace*{-.5mm}D_{1}^{{\rm o}}\,(92.9\%)$             & $2.093231$ & $2.072514$ & $1.985320$ \\
11 & $3s^{2}3p^{5}3d$ & ${}^{3}\hspace*{-.5mm}F_{2}^{{\rm o}}\,(81.3\%)$             & $1.562998$ & $1.563029$ & $1.494070$ &
11 & $3s^{2}3p^{5}3d$ & ${}^{3}\hspace*{-.5mm}D_{2}^{{\rm o}}\,(54.0\%)$             & $2.095232$ & $2.072117$ & $1.984866$ \\
\multicolumn{12}{c}{\vspace*{-2mm}} \\
12 & $3s^{2}3p^{5}3d$ & ${}^{3}\hspace*{-.5mm}D_{3}^{{\rm o}}\,(35.7\%)$             & $1.633040$ & $1.634775$ & $1.561671$ &
12 & $3s^{2}3p^{5}3d$ & ${}^{1}\hspace*{-.5mm}F_{3}^{{\rm o}}\,(57.8\%)$             & $2.081461$ & $2.097941$ & $1.995780$ \\
\multicolumn{12}{c}{\vspace*{-2mm}} \\
13 & $3s^{2}3p^{5}3d$ & ${}^{1}\hspace*{-.5mm}D_{2}^{{\rm o}}\,(41.6\%)$             & $1.639503$ & $1.634350$ & $1.560836$ &
13 & $3s^{2}3p^{5}4s$ & ${}^{3}\hspace*{-.5mm}P_{2}^{{\rm o}}\,(99.4\%)$             & $2.313063$ & $2.210253$ & $2.114045$ \\
\multicolumn{12}{c}{\vspace*{-2mm}} \\
14 & $3s^{2}3p^{5}3d$ & ${}^{3}\hspace*{-.5mm}D_{1}^{{\rm o}}\,(71.5\%)$                       & $1.644348$ & $1.644371$ & $1.570061$ &
14 & $3s^{2}3p^{5}4s$ & ${}^{3}\hspace*{-.5mm}P_{1}^{{\rm o}}\,(88.0\%)$                       & $2.327118$ & $2.222858$ & $2.126335$ \\
15 & $3s^{2}3p^{5}3d$ & ${}^{3}\hspace*{-.5mm}D_{2}^{{\rm o}}\,(41.9\%)$                       & $1.645874$ & $1.643859$ & $1.569134$ &
15 & $3s^{2}3p^{5}4s$ & ${}^{3}\hspace*{-.5mm}P_{0}^{{\rm o}}\,(99.4\%)$                       & $2.341390$ & $2.238180$ & $2.140287$ \\
16 & $3s^{2}3p^{5}3d$ & ${}^{1}\hspace*{-.5mm}F_{3}^{{\rm o}}\,(34.1\%)$                       & $1.645999$ & $1.645919$ & $1.571404$ &
16 & $3s^{2}3p^{5}4s$ & ${}^{1}\hspace*{-.5mm}P_{1}^{{\rm o}}\,(87.6\%)$                       & $2.369020$ & $2.257176$ & $2.160581$ \\
\multicolumn{12}{c}{\vspace*{-2mm}} \\
17 & $3s^{2}3p^{5}4p$ & ${}^{3}\hspace*{-.5mm}S_{1}\,(97.7\%)$                                    & $1.663215$ & $1.669479$ & $1.591351$ &
17 & $3s^{2}3p^{5}3d$ & ${}^{1}\hspace*{-.5mm}P_{1}^{{\rm o}}\,(63.1\%)^{\bigstar}$            & $2.545658$ & $2.545658$ & $2.492690$ \\
\multicolumn{12}{c}{\vspace*{-2mm}} \\
18 & $3s^{2}3p^{5}4p$ & ${}^{3}\hspace*{-.5mm}D_{3}\,(99.9\%)$            & $1.693675$ & $1.698454$ & $1.618318$ &
   &                  &                                                &         &         &             \\
19 & $3s^{2}3p^{5}4p$ & ${}^{3}\hspace*{-.5mm}D_{2}\,(75.5\%)$            & $1.697258$ & $1.701166$ & $1.621064$ &
   &                  &                                                &         &         &             \\
20 & $3s^{2}3p^{5}4p$ & ${}^{3}\hspace*{-.5mm}D_{1}\,(76.3\%)$            & $1.705934$ & $1.708872$ & $1.628229$ &
   &                  &                                                &         &         &             \\
21 & $3s^{2}3p^{5}4p$ & ${}^{1}\hspace*{-.5mm}D_{2}\,(51.0\%)$            & $1.712827$ & $1.714552$ & $1.633790$ &
   &                  &                                                &         &         &             \\
22 & $3s^{2}3p^{5}4p$ & ${}^{1}\hspace*{-.5mm}P_{1}\,(52.5\%)$            & $1.724516$ & $1.724477$ & $1.643195$ &
   &                  &                                                &         &         &             \\
23 & $3s^{2}3p^{5}4p$ & ${}^{3}\hspace*{-.5mm}P_{2}\,(55.9\%)$            & $1.727938$ & $1.728285$ & $1.646621$ &
   &                  &                                                &         &         &             \\
24 & $3s^{2}3p^{5}4p$ & ${}^{3}\hspace*{-.5mm}P_{0}\,(99.5\%)$            & $1.730260$ & $1.729293$ & $1.648054$ &
   &                  &                                                &         &         &             \\
25 & $3s^{2}3p^{5}4p$ & ${}^{3}\hspace*{-.5mm}P_{1}\,(65.5\%)$            & $1.733330$ & $1.732598$ & $1.650821$ &
   &                  &                                                &         &         &             \\
26 & $3s^{2}3p^{5}4p$ & ${}^{1}\hspace*{-.5mm}S_{0}\,(95.2\%)$            & $1.829238$ & $1.774895$ & $1.702454$ &
   &                  &                                                &         &         &             \\
27 & $3s^{2}3p^{5}3d$ & ${}^{1}\hspace*{-.5mm}P_{1}^{{\rm o}}\,(90.5\%)^{\bigstar}$ & $1.840715$ & $1.840373$ & $1.817999$ &
   &                  &                                                          &         &         &   \\
   & $3s^{2}3p^{5}4d$ & ${}^{1}\hspace*{-.5mm}P_{1}^{{\rm o}}\,(90.0\%)$            & $2.033261$ & $2.033256$ &  &
   &                  &                                                &         &         &             \\
\hline
\end{tabular}
} 

    \begin{list}{}{}
    \item[$^{{\rm a}}$] present work: 51-level (${\rm K}^{+}$) and 69-level (${\rm Ca}^{2+}$) MCBP results
    \item[$^{{\rm b}}$] UV spark spectroscopy experimental data of \cite{Pettersen:2007}
    \item[$^{{\rm c}}$] MCHF results of \cite{Irimia:2003}
    \item[$^{{\rm d}}$] critically compiled experimental data of \cite{Sugar:1985}
    \item[$^{\bigstar}  $] dominant excitation threshold - see Table~\ref{Tab:tar-rad}
    \end{list}
\end{minipage}
\end{table*}

\clearpage

\begin{table*}[!hbtp]
\begin{minipage}[t]{\textwidth}
\caption{\label{Tab:V:Cr:tar-str}
The lowest $\Delta n_{{\rm c}} = 0$ core excitation thresholds (in Rydbergs)
for \ion{V}{vi} and \ion{Cr}{vii}.
}
\resizebox{\textwidth}{!}{
\begin{tabular}{r l l l r r c l l r r}  
\hline
\hline
\multicolumn{11}{c}{\vspace*{-2mm}} \\
   &         & \multicolumn{4}{c}{${\rm V}^{5+}$} &   & \multicolumn{4}{c}{${\rm Cr}^{6+}$} \\
\multicolumn{11}{c}{\vspace*{-2mm}} \\
\cline{3-6}\cline{8-11}
\multicolumn{11}{c}{\vspace*{-2mm}} \\
K  & Config. & Level(mix) & Present$^{{\rm a}}$ & NIST$^{{\rm b}}$ & MCHF$^{{\rm c}}$ &
             & Level(mix) & Present$^{{\rm a}}$ & NIST$^{{\rm b}}$ & MCHF$^{{\rm c}}$ \\
\hline
\multicolumn{11}{c}{\vspace*{-2mm}} \\
 1 & $3s^{2}3p^{6}  $ & ${}^{1}\hspace*{-.5mm}S_{0}\,(96.4\%)$                       & $0.000000$ & $0.000000$ & $0.000000$ &
                      & ${}^{1}\hspace*{-.5mm}S_{0}\,(96.6\%)$                       & $0.000000$ & $0.000000$ & $0.000000$ \\
\multicolumn{11}{c}{\vspace*{-2mm}} \\
 2 & $3s^{2}3p^{5}3d$ & ${}^{3}\hspace*{-.5mm}P_{0}^{{\rm o}}\,(97.0\%)$             & $2.739060$ & $2.808068$ & $2.806866$ &
                      & ${}^{3}\hspace*{-.5mm}P_{0}^{{\rm o}}\,(97.1\%)$             & $3.059906$ & $3.109055$ & $3.110232$ \\
 3 & $3s^{2}3p^{5}3d$ & ${}^{3}\hspace*{-.5mm}P_{1}^{{\rm o}}\,(96.9\%)$             & $2.748293$ & $2.819413$ & $2.818090$ &
                      & ${}^{3}\hspace*{-.5mm}P_{1}^{{\rm o}}\,(96.9\%)$             & $3.072680$ & $3.123582$ & $3.124881$ \\
 4 & $3s^{2}3p^{5}3d$ & ${}^{3}\hspace*{-.5mm}P_{2}^{{\rm o}}\,(96.7\%)$             & $2.767020$ & $2.842952$ & $2.841350$ &
                      & ${}^{3}\hspace*{-.5mm}P_{2}^{{\rm o}}\,(96.6\%)$             & $3.098672$ & $3.154233$ & $3.155270$ \\
\multicolumn{11}{c}{\vspace*{-2mm}} \\
 5 & $3s^{2}3p^{5}3d$ & ${}^{3}\hspace*{-.5mm}F_{4}^{{\rm o}}\,(97.3\%)$             & $2.896923$ & $2.941329$ & $2.944497$ &
                      & ${}^{3}\hspace*{-.5mm}F_{4}^{{\rm o}}\,(97.4\%)$             & $3.234927$ & $3.258178$ & $3.264649$ \\
 6 & $3s^{2}3p^{5}3d$ & ${}^{3}\hspace*{-.5mm}F_{3}^{{\rm o}}\,(96.2\%)$             & $2.906694$ & $2.961235$ & $2.963277$ &
                      & ${}^{3}\hspace*{-.5mm}F_{3}^{{\rm o}}\,(95.7\%)$             & $3.247762$ & $3.282128$ & $3.287197$ \\
 7 & $3s^{2}3p^{5}3d$ & ${}^{3}\hspace*{-.5mm}F_{2}^{{\rm o}}\,(96.6\%)$             & $2.917192$ & $2.981802$ & $2.982698$ &
                      & ${}^{3}\hspace*{-.5mm}F_{2}^{{\rm o}}\,(96.2\%)$             & $3.262960$ & $3.308454$ & $3.312214$ \\
\multicolumn{11}{c}{\vspace*{-2mm}} \\
 8 & $3s^{2}3p^{5}3d$ & ${}^{1}\hspace*{-.5mm}D_{2}^{{\rm o}}\,(71.5\%)$             & $3.110770$ & $3.145142$ & $3.152963$ &
                      & ${}^{1}\hspace*{-.5mm}D_{2}^{{\rm o}}\,(67.4\%)$             & $3.470531$ & $3.487258$ & $3.497713$ \\
\multicolumn{11}{c}{\vspace*{-2mm}} \\
 9 & $3s^{2}3p^{5}3d$ & ${}^{3}\hspace*{-.5mm}D_{3}^{{\rm o}}\,(78.2\%)$             & $3.119456$ & $3.148578$ & $3.156780$ &
                      & ${}^{3}\hspace*{-.5mm}D_{3}^{{\rm o}}\,(72.3\%)$             & $3.478260$ & $3.487760$ & $3.499169$ \\
10 & $3s^{2}3p^{5}3d$ & ${}^{3}\hspace*{-.5mm}D_{1}^{{\rm o}}\,(96.8\%)$             & $3.126389$ & $3.170297$ & $3.176499$ &
                      & ${}^{3}\hspace*{-.5mm}D_{1}^{{\rm o}}\,(96.8\%)$             & $3.489568$ & $3.515926$ & $3.525054$ \\
11 & $3s^{2}3p^{5}3d$ & ${}^{3}\hspace*{-.5mm}D_{2}^{{\rm o}}\,(71.6\%)$             & $3.132258$ & $3.174174$ & $3.180891$ &
                      & ${}^{3}\hspace*{-.5mm}D_{2}^{{\rm o}}\,(71.8\%)$             & $3.499112$ & $3.523110$ & $3.532855$ \\
\multicolumn{11}{c}{\vspace*{-2mm}} \\
12 & $3s^{2}3p^{5}3d$ & ${}^{1}\hspace*{-.5mm}F_{3}^{{\rm o}}\,(77.6\%)$             & $3.145710$ & $3.195308$ & $3.203371$ &
                      & ${}^{1}\hspace*{-.5mm}F_{3}^{{\rm o}}\,(71.3\%)$             & $3.512852$ & $3.546890$ & $3.557753$ \\
\multicolumn{11}{c}{\vspace*{-2mm}} \\
13 & $3s^{2}3p^{5}3d$ & ${}^{1}\hspace*{-.5mm}P_{1}^{{\rm o}}\,(95.4\%)^{\bigstar}$  & $4.059132$ & $4.059108$ & $4.114305$ &
                      & ${}^{1}\hspace*{-.5mm}P_{1}^{{\rm o}}\,(95.6\%)^{\bigstar}$  & $4.492846$ & $4.492869$ & $4.553079$ \\
\multicolumn{11}{c}{\vspace*{-2mm}} \\
14 & $3s^{ }3p^{6}3d$ & ${}^{3}\hspace*{-.5mm}D_{1}\,(69.1\%)$                       & $5.013754$ & $5.007759$ & $5.122273$ &
                      & ${}^{3}\hspace*{-.5mm}D_{1}\,(71.9\%)$                       & $5.547059$ & $5.546697$ & $5.667028$ \\
15 & $3s^{ }3p^{6}3d$ & ${}^{3}\hspace*{-.5mm}D_{2}\,(69.1\%)$                       & $5.017989$ & $5.010726$ & $5.124985$ &
                      & ${}^{3}\hspace*{-.5mm}D_{2}\,(71.8\%)$                       & $5.552838$ & $5.550917$ & $5.670942$ \\
16 & $3s^{ }3p^{6}3d$ & ${}^{3}\hspace*{-.5mm}D_{3}\,(69.2\%)$                       & $5.024692$ & $5.015474$ & $5.129131$ &
                      & ${}^{3}\hspace*{-.5mm}D_{3}\,(72.0\%)$                       & $5.562141$ & $5.557707$ & $5.676969$ \\
\multicolumn{11}{c}{\vspace*{-2mm}} \\
17 & $3s^{ }3p^{6}3d$ & ${}^{1}\hspace*{-.5mm}D_{2}\,(63.6\%)$                       & $5.191350$ & $5.161717$ & $5.318685$ &
                      & ${}^{1}\hspace*{-.5mm}D_{2}\,(66.4\%)$                       & $5.742414$ & $5.721178$ & $5.883764$ \\
\hline
\end{tabular}
} 

    \begin{list}{}{}
    \item[$^{{\rm a}}$] present work: 2894-level MCBP results\
    \item[$^{{\rm b}}$] critically compiled experimental data of \cite{Shirai:2000}
    \item[$^{{\rm c}}$]   MCHF results of \cite{Irimia:2003}
    \item[$^{\bigstar}  $] dominant excitation threshold - see Table~\ref{Tab:tar-rad}
    \end{list}
\end{minipage}
\end{table*}

\clearpage
\begin{table*}[!hbtp]
\begin{minipage}[t]{\textwidth}
\caption{\label{Tab:Mn:Fe:tar-str}
The lowest $\Delta n_{{\rm c}} = 0$ core excitation thresholds (in Rydbergs) for \ion{Mn}{viii} and \ion{Fe}{ix}.
}
\resizebox{\textwidth}{!}{
\begin{tabular}{r l l l r r c l l r r r r r}  
\hline
\hline
\multicolumn{14}{c}{\vspace*{-2mm}} \\
   &         & \multicolumn{4}{c}{${\rm Mn}^{7+}$} &   & \multicolumn{7}{c}{${\rm Fe}^{8+}$} \\
\multicolumn{14}{c}{\vspace*{-2mm}} \\
\cline{3-6}\cline{8-14}
\multicolumn{11}{c}{\vspace*{-2mm}} \\
K  & Config. & Level(mix) & Present$^{{\rm a}}$ & NIST$^{{\rm b}}$ & MCHF$^{{\rm c}}$ &
             & Level(mix) & Present$^{{\rm a}}$ & NIST$^{{\rm b}}$ & MCHF$^{{\rm c}}$ & CIV3$^{{\rm e}}$ & IRON$^{{\rm f}}$ & FAC3$^{{\rm g}}$ \\
\hline
\multicolumn{14}{c}{\vspace*{-2mm}} \\
 1 & $3s^{2}3p^{6}  $ & ${}^{1}\hspace*{-.5mm}S_{0}\,(96.7\%)          $            & $0.00000$ & $0.00000$  & $0.00000$ &
                      & ${}^{1}\hspace*{-.5mm}S_{0}\,(96.9\%)          $            & $0.00000$ & $0.00000$  & $0.00000$ & $0.00000$ & $0.00000$ & $0.0000$ \\
\multicolumn{14}{c}{\vspace*{-2mm}} \\
 2 & $3s^{2}3p^{5}3d$ & ${}^{3}\hspace*{-.5mm}P_{0}^{{\rm o}}\,(97.2\%)$            & $3.36224$ & $3.40502$  & $3.19510$ &
                      & ${}^{3}\hspace*{-.5mm}P_{0}^{{\rm o}}\,(97.4\%)$            & $3.67368$ & $3.69767$  & $3.45516$ & $3.69766$ & $3.76364$ & $3.7475$ \\
 3 & $3s^{2}3p^{5}3d$ & ${}^{3}\hspace*{-.5mm}P_{1}^{{\rm o}}\,(96.9\%)$            & $3.37921$ & $3.42372$  & $3.21381$ &
                      & ${}^{3}\hspace*{-.5mm}P_{1}^{{\rm o}}\,(96.9\%)$            & $3.69563$ & $3.72084$  & $3.47859$ & $3.72078$ & $3.78883$ & $3.7710$ \\
 4 & $3s^{2}3p^{5}3d$ & ${}^{3}\hspace*{-.5mm}P_{2}^{{\rm o}}\,(96.4\%)$            & $3.41384$ & $3.46275$  & $3.25264$ &
                      & ${}^{3}\hspace*{-.5mm}P_{2}^{{\rm o}}\,(96.2\%)$            & $3.74058$ & $3.76963$  & $3.52725$ & $3.76962$ & $3.84039$ & $3.8200$ \\
\multicolumn{14}{c}{\vspace*{-2mm}} \\
 5 & $3s^{2}3p^{5}3d$ & ${}^{3}\hspace*{-.5mm}F_{4}^{{\rm o}}\,(97.5\%)$            & $3.55446$ & $3.57067$  & $3.36662$ &
                      & ${}^{3}\hspace*{-.5mm}F_{4}^{{\rm o}}\,(97.6\%)$            & $3.88447$ & $3.88026$  & $3.64440$ & $3.88027$ & $3.97113$ & $3.9468$ \\
 6 & $3s^{2}3p^{5}3d$ & ${}^{3}\hspace*{-.5mm}F_{3}^{{\rm o}}\,(94.9\%)$            & $3.57113$ & $3.59878$  & $3.39298$ &
                      & ${}^{3}\hspace*{-.5mm}F_{3}^{{\rm o}}\,(93.9\%)$            & $3.90343$ & $3.91217$  & $3.67440$ & $3.91217$ & $4.00201$ & $3.9797$ \\
 7 & $3s^{2}3p^{5}3d$ & ${}^{3}\hspace*{-.5mm}F_{2}^{{\rm o}}\,(95.5\%)$            & $3.59269$ & $3.63198$  & $3.42463$ &
                      & ${}^{3}\hspace*{-.5mm}F_{2}^{{\rm o}}\,(94.6\%)$            & $3.93169$ & $3.95325$  & $3.71373$ & $3.95329$ & $4.04308$ & $4.0215$ \\
\multicolumn{14}{c}{\vspace*{-2mm}} \\
 8 & $3s^{2}3p^{5}3d$ & ${}^{3}\hspace*{-.5mm}D_{3}^{{\rm o}}\,(68.7\%)$            & $3.81841$ & $3.82162$  & $3.62227$ &
                      & ${}^{3}\hspace*{-.5mm}D_{3}^{{\rm o}}\,(64.5\%)$            & $4.16736$ & $4.15184$  & $3.92030$ & $4.15171$ & $4.25326$ & $4.2373$ \\
\multicolumn{14}{c}{\vspace*{-2mm}} \\
 9 & $3s^{2}3p^{5}3d$ & ${}^{1}\hspace*{-.5mm}D_{2}^{{\rm o}}\,(64.4\%)$            & $3.81303$ & $3.82565$  & $3.62467$ &
                      & ${}^{1}\hspace*{-.5mm}D_{2}^{{\rm o}}\,(62.1\%)$            & $4.16451$ & $4.16224$  & $3.92846$ & $4.16220$ & $4.26847$ & $4.2540$ \\
\multicolumn{14}{c}{\vspace*{-2mm}} \\
10 & $3s^{2}3p^{5}3d$ & ${}^{3}\hspace*{-.5mm}D_{1}^{{\rm o}}\,(96.8\%)$            & $3.83602$ & $3.85773$  & $3.65585$ &
                      & ${}^{3}\hspace*{-.5mm}D_{1}^{{\rm o}}\,(96.8\%)$            & $4.19143$ & $4.19744$  & $3.96326$ & $4.19745$ & $4.29702$ & $4.2836$ \\
11 & $3s^{2}3p^{5}3d$ & ${}^{3}\hspace*{-.5mm}D_{2}^{{\rm o}}\,(64.7\%)$            & $3.85013$ & $3.86961$  & $3.66842$ &
                      & ${}^{3}\hspace*{-.5mm}D_{2}^{{\rm o}}\,(62.6\%)$            & $4.21227$ & $4.21567$  & $3.98226$ & $4.21573$ & $4.31956$ & $4.3039$ \\
\multicolumn{14}{c}{\vspace*{-2mm}} \\
12 & $3s^{2}3p^{5}3d$ & ${}^{1}\hspace*{-.5mm}F_{3}^{{\rm o}}\,(67.1\%)$            & $3.86567$ & $3.89595$  & $3.69578$ &
                      & ${}^{1}\hspace*{-.5mm}F_{3}^{{\rm o}}\,(62.2\%)$            & $4.22812$ & $4.24494$  & $4.01241$ & $4.24509$ & $4.34420$ & $4.3281$ \\
\multicolumn{14}{c}{\vspace*{-2mm}} \\
13 & $3s^{2}3p^{5}3d$ & ${}^{1}\hspace*{-.5mm}P_{1}^{{\rm o}}\,(95.7\%)^{\bigstar}$ & $4.91368$ & $4.91368$  & $4.76446$ &
                      & ${}^{1}\hspace*{-.5mm}P_{1}^{{\rm o}}\,(95.9\%)^{\bigstar}$ & $5.32660$ & $5.32678$  & $5.14621$ & $5.32645$ & $5.47936$ & $5.4598$ \\
   &                  &                                                             &           &            &           &
                      &                                                             &           &            & $5.212^{{\rm d}}$ &   &           &          \\
\multicolumn{14}{c}{\vspace*{-2mm}} \\
14 & $3s^{ }3p^{6}3d$ & ${}^{3}\hspace*{-.5mm}D_{1}\,(73.5\%)$                      & $6.05426$ & $6.08432$  & $5.99785$ &
                      & ${}^{3}\hspace*{-.5mm}D_{1}\,(74.7\%)$                      & $6.61242$ & $6.62249$  & $6.50906$ & $6.62249$ & $6.75891$ & $6.7176$ \\
15 & $3s^{ }3p^{6}3d$ & ${}^{3}\hspace*{-.5mm}D_{2}\,(73.4\%)$                      & $6.06173$ & $6.09007$  & $6.00326$ &
                      & ${}^{3}\hspace*{-.5mm}D_{2}\,(74.6\%)$                      & $6.62228$ & $6.63001$  & $6.51629$ & $6.63001$ & $6.76762$ & $6.7251$ \\
16 & $3s^{ }3p^{6}3d$ & ${}^{3}\hspace*{-.5mm}D_{3}\,(73.7\%)$                      & $6.07394$ & $6.09935$  & $6.01167$ &
                      & ${}^{3}\hspace*{-.5mm}D_{3}\,(74.9\%)$                      & $6.63882$ & $6.64254$  & $6.52770$ & $6.64254$ & $6.78194$ & $6.7377$ \\
\multicolumn{14}{c}{\vspace*{-2mm}} \\
17 & $3s^{ }3p^{6}3d$ & ${}^{1}\hspace*{-.5mm}D_{2}\,(68.2\%)$                      & $6.26561$ & $6.27726$  & $6.23324$ &
                      & ${}^{1}\hspace*{-.5mm}D_{2}\,(69.4\%)$                      & $6.84122$ & $6.83333$  & $6.76228$ & $6.83334$ & $6.98165$ & $6.9447$ \\
\hline
\end{tabular}
} 

    \begin{list}{}{}
        \item[$^{{\rm a}} $] present work: 2894-level MCBP results\
        \item[$^{{\rm b}} $] critically compiled experimental data of \cite{Shirai:2000}.
        \item[$^{{\rm c}} $]   MCHF results\ of \cite{Irimia:2003}.
        \item[$^{\bigstar}$] dominant excitation threshold - see Table~\ref{Tab:tar-rad}.
        \item[$^{{\rm d}} $] single configuration TDCHF results\ of \cite{Ghosh:1997}.
        \item[$^{{\rm e}} $] CIV3 results\ of \cite{Verma:2006}.
        \item[$^{{\rm f}} $] IRON Project results\ of \cite{Storey:2002}.
        \item[$^{{\rm g}} $] 6284-level FAC results\ of \cite{Aggarwal:2006}.
    \end{list}
\end{minipage}
\end{table*}

\clearpage
\begin{table*}[!hbtp]
\begin{minipage}[t]{\textwidth}
\caption{\label{Tab:Co:Ni:tar-str}
The lowest $\Delta n_{{\rm c}} = 0$ core excitation thresholds (in Rydbergs) for \ion{Co}{x} and \ion{Ni}{xi}.
Uncertainties are enclosed in lower parentheses.
}
\resizebox{\textwidth}{!}{
\begin{tabular}{r l l l r r c l l r r r r}  
\hline
\hline
\multicolumn{13}{c}{\vspace*{-2mm}} \\
   &         & \multicolumn{4}{c}{${\rm Co}^{9+}$} &   & \multicolumn{6}{c}{${\rm Ni}^{10+}$} \\
\multicolumn{13}{c}{\vspace*{-2mm}} \\
\cline{3-6}\cline{8-13}
\multicolumn{13}{c}{\vspace*{-2mm}} \\
K  & Config. & Level(mix) & Present$^{{\rm a}}$ & NIST$^{{\rm b}}$ & MCHF$^{{\rm c}}$ &
             & Level(mix) & Present$^{{\rm a}}$ & NIST$^{{\rm b}}$ & MCHF$^{{\rm c}}$ & FAC3$^{{\rm e}}$ & CIV3$^{{\rm f}}$ \\
\hline
\multicolumn{13}{c}{\vspace*{-2mm}} \\
 1 & $3s^{2}3p^{6}  $ & ${}^{1}\hspace*{-.5mm}S_{0}\,(97.0\%)          $               & $0.00000$ & $0.00000            $ & $0.00000$ &
                      & ${}^{1}\hspace*{-.5mm}S_{0}\,(97.1\%)          $               & $0.00000$ & $0.00000            $ & $0.00000$ & $0.0000 $ & $0.00000$ \\
\multicolumn{13}{c}{\vspace*{-2mm}} \\
 2 & $3s^{2}3p^{5}3d$ & ${}^{3}\hspace*{-.5mm}P_{0}^{{\rm o}}\,(97.5\%)$               & $3.97623$ & $3.999_{(30)}^{\dag}$ & $3.70893$ &
                      & ${}^{3}\hspace*{-.5mm}P_{0}^{{\rm o}}\,(97.6\%)$               & $4.27882$ & $4.27667            $ & $3.95697$ & $4.2998 $ & $4.34736$ \\
 3 & $3s^{2}3p^{5}3d$ & ${}^{3}\hspace*{-.5mm}P_{1}^{{\rm o}}\,(96.9\%)$               & $4.00373$ & $4.026_{(28)}^{\dag}$ & $3.73778$ &
                      & ${}^{3}\hspace*{-.5mm}P_{1}^{{\rm o}}\,(96.8\%)$               & $4.31288$ & $4.31002            $ & $3.99197$ & $4.3349 $ & $4.38366$ \\
 4 & $3s^{2}3p^{5}3d$ & ${}^{3}\hspace*{-.5mm}P_{2}^{{\rm o}}\,(95.8\%)$               & $4.06022$ & $4.084_{(21)}^{\dag}$ & $3.79773$ &
                      & ${}^{3}\hspace*{-.5mm}P_{2}^{{\rm o}}\,(95.2\%)$               & $4.38307$ & $4.38274            $ & $4.06469$ & $4.4086 $ & $4.45723$ \\
\multicolumn{13}{c}{\vspace*{-2mm}} \\
 5 & $3s^{2}3p^{5}3d$ & ${}^{3}\hspace*{-.5mm}F_{4}^{{\rm o}}\,(97.7\%)$               & $4.20594$ & $4.199_{(31)}^{\dag}$ & $3.91676$ &
                      & ${}^{3}\hspace*{-.5mm}F_{4}^{{\rm o}}\,(97.8\%)$               & $4.52944$ & $4.49482            $ & $4.18448$ & $4.5391 $ & $4.64639$ \\
 6 & $3s^{2}3p^{5}3d$ & ${}^{3}\hspace*{-.5mm}F_{3}^{{\rm o}}\,(92.8\%)$               & $4.22715$ & $4.235_{(31)}^{\dag}$ & $3.94997$ &
                      & ${}^{3}\hspace*{-.5mm}F_{3}^{{\rm o}}\,(91.3\%)$               & $4.55153$ & $4.53374            $ & $4.22023$ & $4.5785 $ & $4.68656$ \\
 7 & $3s^{2}3p^{5}3d$ & ${}^{3}\hspace*{-.5mm}F_{2}^{{\rm o}}\,(93.2\%)$               & $4.26350$ & $4.283_{(25)}^{\dag}$ & $3.99798$ &
                      & ${}^{3}\hspace*{-.5mm}F_{2}^{{\rm o}}\,(91.3\%)$               & $4.59721$ & $4.59342            $ & $4.27778$ & $4.6393 $ & $4.73636$ \\
\multicolumn{13}{c}{\vspace*{-2mm}} \\
 8 & $3s^{2}3p^{5}3d$ & ${}^{3}\hspace*{-.5mm}D_{3}^{{\rm o}}\,(62.3\%)$               & $4.50743$ & $4.494_{(38)}^{\dag}$ & $4.21215$ &
                      & ${}^{3}\hspace*{-.5mm}D_{3}^{{\rm o}}\,(60.3\%)$               & $4.84888$ & $4.80666            $ & $4.49868$ & $4.8593 $ & $4.93579$ \\
\multicolumn{13}{c}{\vspace*{-2mm}} \\
 9 & $3s^{2}3p^{5}3d$ & ${}^{1}\hspace*{-.5mm}D_{2}^{{\rm o}}\,(60.5\%)$               & $4.50846$ & $4.511_{(29)}^{\dag}$ & $4.22855$ &
                      & ${}^{1}\hspace*{-.5mm}D_{2}^{{\rm o}}\,(58.8\%)$               & $4.85545$ & $4.83728            $ & $4.52643$ & $4.8953 $ & $4.97061$ \\
\multicolumn{13}{c}{\vspace*{-2mm}} \\
10 & $3s^{2}3p^{5}3d$ & ${}^{3}\hspace*{-.5mm}D_{1}^{{\rm o}}\,(96.7\%)$               & $4.53943$ & $4.53611            $ & $4.26634$ &
                      & ${}^{3}\hspace*{-.5mm}D_{1}^{{\rm o}}\,(96.5\%)$               & $4.88993$ & $4.87373            $ & $4.56612$ & $4.9290 $ & $5.02599$ \\
11 & $3s^{2}3p^{5}3d$ & ${}^{3}\hspace*{-.5mm}D_{2}^{{\rm o}}\,(61.4\%)$               & $4.56821$ & $4.573_{(24)}^{\dag}$ & $4.29382$ &
                      & ${}^{3}\hspace*{-.5mm}D_{2}^{{\rm o}}\,(60.2\%)$               & $4.92948$ & $4.91337            $ & $4.60454$ & $4.9687 $ & $5.04846$ \\
\multicolumn{13}{c}{\vspace*{-2mm}} \\
12 & $3s^{2}3p^{5}3d$ & ${}^{1}\hspace*{-.5mm}F_{3}^{{\rm o}}\,(59.0\%)$               & $4.58588$ & $4.606_{(25)}^{\dag}$ & $4.32726$ &
                      & ${}^{1}\hspace*{-.5mm}F_{3}^{{\rm o}}\,(56.1\%)$               & $4.94891$ & $4.95018            $ & $4.64195$ & $5.0024 $ & $5.12673$ \\
\multicolumn{13}{c}{\vspace*{-2mm}} \\
13 & $3s^{2}3p^{5}3d$ & ${}^{1}\hspace*{-.5mm}P_{1}^{{\rm o}}\,(96.1\%)^{\bigstar}$    & $5.73579$ & $5.73579            $ & $5.51933$ &
                      & ${}^{1}\hspace*{-.5mm}P_{1}^{{\rm o}}\,(96.2\%)^{\bigstar}$    & $6.14044$ & $6.14158            $ & $5.88618$ & $6.2574 $ & $6.41884$ \\
   &                  &                                                                &           &                       & $5.5952^{{\rm d}}$ &
                      &                                                                &           &                       & $5.9728^{{\rm d}}$ &  &           \\
\multicolumn{13}{c}{\vspace*{-2mm}} \\
14 & $3s^{ }3p^{6}3d$ & ${}^{3}\hspace*{-.5mm}D_{1}\,(74.2\%)          $               & $7.21441$ & $7.178_{(23)}^{\dag}$ & $7.01984$ &
                      & ${}^{3}\hspace*{-.5mm}D_{1}\,(75.1\%)          $               & $7.76928$ & $7.720_{(31)}^{\dag}$ & $7.53153$ & $7.7562 $ & $7.85347$ \\
15 & $3s^{ }3p^{6}3d$ & ${}^{3}\hspace*{-.5mm}D_{2}\,(74.0\%)          $               & $7.22694$ & $7.186_{(31)}^{\dag}$ & $7.02926$ &
                      & ${}^{3}\hspace*{-.5mm}D_{2}\,(74.1\%)          $               & $7.78502$ & $7.730_{(32)}^{\dag}$ & $7.54354$ & $7.7687 $ & $7.86852$ \\
16 & $3s^{ }3p^{6}3d$ & ${}^{3}\hspace*{-.5mm}D_{3}\,(74.4\%)          $               & $7.24842$ & $7.198_{(31)}^{\dag}$ & $7.04436$ &
                      & ${}^{3}\hspace*{-.5mm}D_{3}\,(75.3\%)          $               & $7.81290$ & $7.740_{(33)}^{\dag}$ & $7.56317$ & $7.7904 $ & $7.89090$ \\
\multicolumn{13}{c}{\vspace*{-2mm}} \\
17 & $3s^{ }3p^{6}3d$ & ${}^{1}\hspace*{-.5mm}D_{2}\,(68.8\%)          $               & $7.46032$ & $7.410_{(33)}^{\dag}$ & $7.29060$ &
                      & ${}^{1}\hspace*{-.5mm}D_{2}\,(69.7\%)          $               & $8.03277$ & $7.970_{(34)}^{\dag}$ & $7.81976$ & $7.9965 $ & $9.00447$ \\
\hline
\end{tabular}
} 

    \begin{list}{}{}
    \item[$^{{\rm a}} $] present work: 2894-level MCBP results\
    \item[$^{{\rm b}} $] critically compiled experimental data of \cite{Shirai:2000}.
    \item[$^{{\rm c}} $] MCHF results of \cite{Irimia:2003}.
    \item[$^{\dag}    $] extrapolated along the isoelectronic sequence.
    \item[$^{\bigstar}$] dominant excitation threshold - see Table~\ref{Tab:tar-rad}.
    \item[$^{{\rm d}} $] single-configuration TDCHF results\ of \cite{Ghosh:1997}.
    \item[$^{{\rm e}} $] 6164-level FAC results\ of \cite{Aggarwal:2008}.
    \item[$^{{\rm f}} $] restricted CIV3 results\ of \cite{Verma:2007}.
    \end{list}
\end{minipage}
\end{table*}

\clearpage
\begin{table*}[!hbtp]
\begin{minipage}[t]{\textwidth}
\caption{\label{Tab:Cu:Zn:tar-str}
The lowest $\Delta n_{{\rm c}} = 0$ core excitation thresholds (in Rydbergs)
for \ion{Cu}{xii} and \ion{Zn}{xiii}. Uncertainties are enclosed in lower parentheses.
}
\resizebox{\textwidth}{!}{
\begin{tabular}{r l l l r r c l l r r}  
\hline
\hline
\multicolumn{11}{c}{\vspace*{-2mm}} \\
   &         & \multicolumn{4}{c}{${\rm Cu}^{11+}$} &   & \multicolumn{4}{c}{${\rm Zn}^{12+}$} \\
\multicolumn{11}{c}{\vspace*{-2mm}} \\
\cline{3-6}\cline{8-11}
\multicolumn{11}{c}{\vspace*{-2mm}} \\
K  & Config. & Level(mix) & Present$^{{\rm a}}$ & NIST$^{{\rm b}}$ & MCHF$^{{\rm c}}$ &
             & Level(mix) & Present$^{{\rm a}}$ & NIST$^{{\rm b}}$ & MCHF$^{{\rm c}}$ \\
\hline
\multicolumn{11}{c}{\vspace*{-2mm}} \\
 1 & $3s^{2}3p^{6}  $ & ${}^{1}\hspace*{-.5mm}S_{0}\,(97.2\%)$                       & $0.000000$ & $0.000000         $ & $0.000000       $ &
                      & ${}^{1}\hspace*{-.5mm}S_{0}\,(97.3\%)$                       & $0.00000 $ & $0.00000          $ & $0.00000        $ \\
\multicolumn{11}{c}{\vspace*{-2mm}} \\
 2 & $3s^{2}3p^{5}3d$ & ${}^{3}\hspace*{-.5mm}P_{0}^{{\rm o}}\,(97.7\%)$             & $4.56676 $ & $4.603_{(36)}^{\S}$ & $4.19956        $ &
                      & ${}^{3}\hspace*{-.5mm}P_{0}^{{\rm o}}\,(97.7\%)$             & $4.85185 $ & $4.905_{(38)}^{\S}$ & $4.43697        $ \\
 3 & $3s^{2}3p^{5}3d$ & ${}^{3}\hspace*{-.5mm}P_{1}^{{\rm o}}\,(96.6\%)$             & $4.60819 $ & $4.638_{(32)}^{\S}$ & $4.24146        $ &
                      & ${}^{3}\hspace*{-.5mm}P_{1}^{{\rm o}}\,(96.3\%)$             & $4.90147 $ & $4.943_{(35)}^{\S}$ & $4.48650        $ \\
 4 & $3s^{2}3p^{5}3d$ & ${}^{3}\hspace*{-.5mm}P_{2}^{{\rm o}}\,(94.4\%)$             & $4.69378 $ & $4.711_{(25)}^{\S}$ & $4.32844        $ &
                      & ${}^{3}\hspace*{-.5mm}P_{2}^{{\rm o}}\,(93.4\%)$             & $5.00416 $ & $5.025_{(27)}^{\S}$ & $4.58924        $ \\
\multicolumn{11}{c}{\vspace*{-2mm}} \\
 5 & $3s^{2}3p^{5}3d$ & ${}^{3}\hspace*{-.5mm}F_{4}^{{\rm o}}\,(97.9\%)$             & $4.83899 $ & $4.837_{(36)}^{\S}$ & $4.44807        $ &
                      & ${}^{3}\hspace*{-.5mm}F_{4}^{{\rm o}}\,(97.9\%)$             & $5.14710 $ & $5.156_{(38)}^{\S}$ & $4.70801        $ \\
 6 & $3s^{2}3p^{5}3d$ & ${}^{3}\hspace*{-.5mm}F_{3}^{{\rm o}}\,(89.6\%)$             & $4.86273 $ & $4.880_{(35)}^{\S}$ & $4.48537        $ &
                      & ${}^{3}\hspace*{-.5mm}F_{3}^{{\rm o}}\,(87.7\%)$             & $5.17130 $ & $5.203_{(38)}^{\S}$ & $4.74554        $ \\
 7 & $3s^{2}3p^{5}3d$ & ${}^{3}\hspace*{-.5mm}F_{2}^{{\rm o}}\,(88.5\%)$             & $4.92015 $ & $4.941_{(29)}^{\S}$ & $4.55309        $ &
                      & ${}^{3}\hspace*{-.5mm}F_{2}^{{\rm o}}\,(84.9\%)$             & $5.24156 $ & $5.270_{(31)}^{\S}$ & $4.82380        $ \\
\multicolumn{11}{c}{\vspace*{-2mm}} \\
 8 & $3s^{2}3p^{5}3d$ & ${}^{1}\hspace*{-.5mm}D_{2}^{{\rm o}}\,(57.2\%)$             & $5.19405 $ & $5.203_{(34)}^{\S}$ & $4.82348        $ &
                      & ${}^{1}\hspace*{-.5mm}D_{2}^{{\rm o}}\,(55.4\%)$             & $5.53462 $ & $5.548_{(36)}^{\S}$ & $5.12106        $ \\
\multicolumn{11}{c}{\vspace*{-2mm}} \\
 9 & $3s^{2}3p^{5}3d$ & ${}^{3}\hspace*{-.5mm}D_{3}^{{\rm o}}\,(59.8\%)$             & $5.17774 $ & $5.177_{(44)}^{\S}$ & $4.78044        $ &
                      & ${}^{3}\hspace*{-.5mm}D_{3}^{{\rm o}}\,(59.6\%)$             & $5.50490 $ & $5.519_{(47)}^{\S}$ & $5.05792        $ \\
10 & $3s^{2}3p^{5}3d$ & ${}^{3}\hspace*{-.5mm}D_{1}^{{\rm o}}\,(96.2\%)$             & $5.23130 $ & $5.21500          $ & $4.86330        $ &
                      & ${}^{3}\hspace*{-.5mm}D_{1}^{{\rm o}}\,(95.8\%)$             & $5.57297 $ & $5.55700          $ & $5.15855        $ \\
11 & $3s^{2}3p^{5}3d$ & ${}^{3}\hspace*{-.5mm}D_{2}^{{\rm o}}\,(59.4\%)$             & $5.28354 $ & $5.280_{(32)}^{\S}$ & $4.91565        $ &
                      & ${}^{3}\hspace*{-.5mm}D_{2}^{{\rm o}}\,(58.7\%)$             & $5.64097 $ & $5.633_{(35)}^{\S}$ & $5.22837        $ \\
\multicolumn{11}{c}{\vspace*{-2mm}} \\
12 & $3s^{2}3p^{5}3d$ & ${}^{1}\hspace*{-.5mm}F_{3}^{{\rm o}}\,(54.3\%)$             & $5.30733 $ & $5.320_{(29)}^{\S}$ & $4.95778        $ &
                      & ${}^{1}\hspace*{-.5mm}F_{3}^{{\rm o}}\,(52.7\%)$             & $5.67000 $ & $5.676_{(31)}^{\S}$ & $5.27605        $ \\
\multicolumn{11}{c}{\vspace*{-2mm}} \\
13 & $3s^{2}3p^{5}3d$ & ${}^{1}\hspace*{-.5mm}P_{1}^{{\rm o}}\,(96.2\%)^{\bigstar}$  & $6.54764 $ & $6.54764          $ & $6.24830        $ &
                      & ${}^{1}\hspace*{-.5mm}P_{1}^{{\rm o}}\,(96.1\%)^{\bigstar}$  & $6.95380 $ & $6.95379          $ & $6.60700        $ \\
   &                  &                                                              &            &                     & $6.346 ^{{\rm d}}$&
                      &                                                              &            &                     & $6.7154^{{\rm d}}$\\
\multicolumn{11}{c}{\vspace*{-2mm}} \\
14 & $3s^{ }3p^{6}3d$ & ${}^{3}\hspace*{-.5mm}D_{1}\,(75.6\%)$                       & $8.29610 $ & $8.268_{(27)}^{\S}$ & $8.04518        $ &
                      & ${}^{3}\hspace*{-.5mm}D_{1}\,(75.9\%)$                       & $8.82751 $ & $8.814_{(29)}^{\S}$ & $8.56194        $ \\
15 & $3s^{ }3p^{6}3d$ & ${}^{3}\hspace*{-.5mm}D_{2}\,(74.9\%)$                       & $8.31511 $ & $8.277_{(36)}^{\S}$ & $8.06020        $ &
                      & ${}^{3}\hspace*{-.5mm}D_{2}\,(75.1\%)$                       & $8.85018 $ & $8.822_{(38)}^{\S}$ & $8.58042        $ \\
16 & $3s^{ }3p^{6}3d$ & ${}^{3}\hspace*{-.5mm}D_{3}\,(75.9\%)$                       & $8.34975 $ & $8.291_{(36)}^{\S}$ & $8.08529        $ &
                      & ${}^{3}\hspace*{-.5mm}D_{3}\,(76.3\%)$                       & $8.89283 $ & $8.838_{(39)}^{\S}$ & $8.61207        $ \\
\multicolumn{11}{c}{\vspace*{-2mm}} \\
17 & $3s^{ }3p^{6}3d$ & ${}^{1}\hspace*{-.5mm}D_{2}\,(70.1\%)$                       & $8.57720 $ & $8.538_{(39)}^{\S}$ & $8.35116        $ &
                      & ${}^{1}\hspace*{-.5mm}D_{2}\,(70.3\%)$                       & $9.12726 $ & $9.102_{(42)}^{\S}$ & $8.88624        $ \\
\hline
\end{tabular}
} 

    \begin{list}{}{}
    \item[$^{{\rm a}} $] present work: 2894-level MCBP results
    \item[$^{{\rm b}} $] critically compiled experimental data of \cite{Shirai:2000}
    \item[$^{{\rm c}} $]   MCHF results of \cite{Irimia:2003}
    \item[$^{\S}      $] extrapolated along the isoelectronic sequence
    \item[$^{\bigstar}$] dominant excitation threshold - see Table~\ref{Tab:tar-rad}
    \item[$^{{\rm d}} $] single configuration TDCHF results of \cite{Ghosh:1997}
    \end{list}
\end{minipage}
\end{table*}

\clearpage
\begin{table*}
\begin{minipage}[c]{\textwidth}
\caption{\label{Tab:DR:Maxwl}
Fitting parameters $E_{i}\,({\rm K})$ and $c_{i}\,({\rm cm}^{3} {\rm s}^{-1} {\rm K}^{3/2})$
used for modeling the Maxwellian-averaged DR rate coefficients - see Eq.~(\ref{Eq:Fit:DR}).
Uncertainties are enclosed in lower parentheses, where $v_{(u)}^{[\pm p]}$ denotes $v(u)\times 10^{\pm p}$.
}
\resizebox{\textwidth}{!}{
\begin{tabular}{c l l l l l l l l l l l l l l}   
\hline
\hline
\multicolumn{15}{c}{\vspace*{-2mm}} \\
ion      &
$E_1$                & $E_2$                & $E_3$                & $E_4$                & $E_5$                & $E_6$                & $E_7$    &
$c_1$                & $c_2$                & $c_3$                & $c_4$                & $c_5$                & $c_6$                & $c_7$    \\
\multicolumn{15}{c}{\vspace*{-2mm}} \\
\hline
\multicolumn{15}{c}{\vspace*{-2mm}} \\
\multirow{1}{*}{${\rm K}^{+}$} &
$2.451_{(1)}^{[+5]}$ & $3.504_{(1)}^{[+5]}$ & $4.094_{(2)}^{[+5]}$ & $4.766_{(2)}^{[+5]}$ & $\cdots            $ & $\cdots            $ & $\cdots$ &
$6.292_{(1)}^{[-4]}$ & $2.350_{(1)}^{[-3]}$ & $1.165_{(1)}^{[-2]}$ & $2.470_{(1)}^{[-3]}$ & $\cdots            $ & $\cdots            $ & $\cdots$ \\
\multicolumn{15}{c}{\vspace*{-2mm}} \\
\hline
\multicolumn{15}{c}{\vspace*{-2mm}} \\
\multirow{1}{*}{${\rm Ca}^{2+}$} &
$2.282_{(4)}^{[+5]}$ & $3.682_{(1)}^{[+5]}$ & $4.479_{(4)}^{[+5]}$ & $\cdots            $ & $\cdots            $ & $\cdots            $ & $\cdots$ &
$3.843_{(2)}^{[-4]}$ & $8.040_{(3)}^{[-3]}$ & $8.670_{(5)}^{[-3]}$ & $\cdots            $ & $\cdots            $ & $\cdots            $ & $\cdots$ \\
\multicolumn{15}{c}{\vspace*{-2mm}} \\
\hline
\multicolumn{15}{c}{\vspace*{-2mm}} \\
\multirow{1}{*}{${\rm Sc}^{3+}$} &
$1.168_{(1)}^{[+5]}$ & $2.188_{(2)}^{[+5]}$ & $4.771_{(3)}^{[+5]}$ & $6.814_{(1)}^{[+5]}$ & $\cdots            $ & $\cdots            $ & $\cdots$ &
$9.817_{(2)}^{[-5]}$ & $6.241_{(1)}^{[-4]}$ & $2.459_{(2)}^{[-2]}$ & $6.990_{(5)}^{[-3]}$ & $\cdots            $ & $\cdots            $ & $\cdots$ \\
\multicolumn{15}{c}{\vspace*{-2mm}} \\
\hline
\multicolumn{15}{c}{\vspace*{-2mm}} \\
\multirow{1}{*}{${\rm Ti}^{4+}$} &
$1.5603_{(2)}^{[+1]}$ & $9.8857_{(1)}^{[+1]}$ & $3.7145_{(3)}^{[+2]}$ & $1.4089_{(3)}^{[+3]}$ & $5.9207_{(1)}^{[+3]}$ & $4.1099_{(6)}^{[+4]}$ & $5.1296_{(1)}^{[+5]}$ &
$1.006_{(7)}^{[-6]} $ & $1.2449_{(1)}^{[-6]}$ & $4.3180_{(1)}^{[-6]}$ & $1.8210_{(7)}^{[-5]}$ & $7.9042_{(5)}^{[-5]}$ & $6.040_{(1)}^{[-4]} $ & $8.457_{(2)}^{[-2]} $ \\
\multicolumn{15}{c}{\vspace*{-2mm}} \\
\hline
\multicolumn{15}{c}{\vspace*{-2mm}} \\
\multirow{1}{*}{${\rm V}^{5+}$} &
$1.655_{(1)}^{[+4]}$ & $4.414_{(2)}^{[+4]}$ & $1.088_{(5)}^{[+5]}$ & $2.299_{(1)}^{[+5]}$ & $4.076_{(1)}^{[+5]}$ & $6.415_{(1)}^{[+5]}$ & $7.743_{(2)}^{[+5]}$ &
$1.968_{(1)}^{[-6]}$ & $1.996_{(2)}^{[-5]}$ & $3.053_{(4)}^{[-4]}$ & $2.020_{(5)}^{[-3]}$ & $1.877_{(2)}^{[-2]}$ & $1.031_{(4)}^{[-1]}$ & $1.569_{(3)}^{[-2]}$ \\
\multicolumn{15}{c}{\vspace*{-2mm}} \\
\hline
\multicolumn{15}{c}{\vspace*{-2mm}} \\
\multirow{1}{*}{${\rm Cr}^{6+}$} &
$5.649_{(3)}^{[+3]}$ & $1.664_{(5)}^{[+4]}$ & $5.521_{(1)}^{[+4]}$ & $1.263_{(4)}^{[+5]}$ & $2.463_{(5)}^{[+5]}$ & $5.499_{(3)}^{[+5]}$ & $7.764_{(5)}^{[+5]}$ &
$3.022_{(3)}^{[-6]}$ & $2.606_{(1)}^{[-5]}$ & $4.024_{(3)}^{[-4]}$ & $2.740_{(5)}^{[-3]}$ & $1.102_{(1)}^{[-2]}$ & $5.602_{(1)}^{[-2]}$ & $1.199_{(6)}^{[-1]}$ \\
\multicolumn{15}{c}{\vspace*{-2mm}} \\
\hline
\multicolumn{15}{c}{\vspace*{-2mm}} \\
\multirow{1}{*}{${\rm Mn}^{7+}$} &
$3.284_{(2)}^{[+3]}$ & $1.111_{(4)}^{[+4]}$ & $2.818_{(1)}^{[+4]}$ & $7.017_{(2)}^{[+4]}$ & $1.681_{(3)}^{[+5]}$ & $4.343_{(1)}^{[+5]}$ & $8.056_{(4)}^{[+5]}$ &
$4.596_{(4)}^{[-5]}$ & $1.499_{(2)}^{[-4]}$ & $1.960_{(3)}^{[-3]}$ & $5.060_{(5)}^{[-3]}$ & $7.040_{(5)}^{[-3]}$ & $3.421_{(1)}^{[-2]}$ & $1.963_{(4)}^{[-1]}$ \\
\multicolumn{15}{c}{\vspace*{-2mm}} \\
\hline
\multicolumn{15}{c}{\vspace*{-2mm}} \\
\multirow{1}{*}{${\rm Fe}^{8+}$} &
$2.021_{(3)}^{[+3]}$ & $2.187_{(1)}^{[+3]}$ & $7.063_{(4)}^{[+3]}$ & $2.074_{(1)}^{[+4]}$ & $4.974_{(2)}^{[+4]}$ & $2.743_{(4)}^{[+5]}$ & $8.348_{(1)}^{[+5]}$ &
$3.786_{(4)}^{[-4]}$ & $1.494_{(4)}^{[-4]}$ & $9.447_{(1)}^{[-4]}$ & $2.100_{(5)}^{[-3]}$ & $3.520_{(5)}^{[-3]}$ & $2.785_{(3)}^{[-2]}$ & $2.734_{(5)}^{[-1]}$ \\
\multicolumn{15}{c}{\vspace*{-2mm}} \\
\hline
\multicolumn{15}{c}{\vspace*{-2mm}} \\
\multirow{1}{*}{${\rm Co}^{9+}$} &
$1.277_{(2)}^{[+3]}$ & $9.012_{(4)}^{[+3]}$ & $2.895_{(1)}^{[+4]}$ & $9.202_{(2)}^{[+4]}$ & $1.782_{(3)}^{[+5]}$ & $4.485_{(4)}^{[+5]}$ & $9.033_{(2)}^{[+5]}$ &
$1.549_{(6)}^{[-5]}$ & $2.685_{(1)}^{[-4]}$ & $4.341_{(3)}^{[-4]}$ & $3.670_{(3)}^{[-3]}$ & $1.555_{(2)}^{[-2]}$ & $5.304_{(1)}^{[-2]}$ & $2.857_{(3)}^{[-1]}$ \\
\multicolumn{15}{c}{\vspace*{-2mm}} \\
\hline
\multicolumn{15}{c}{\vspace*{-2mm}} \\
\multirow{1}{*}{${\rm Ni}^{10+}$} &
$1.929_{(2)}^{[+3]}$ & $9.502_{(1)}^{[+3]}$ & $2.690_{(2)}^{[+4]}$ & $6.006_{(2)}^{[+4]}$ & $1.438_{(2)}^{[+5]}$ & $4.627_{(1)}^{[+5]}$ & $1.020_{(2)}^{[+6]}$ &
$1.715_{(1)}^{[-4]}$ & $5.967_{(4)}^{[-4]}$ & $2.660_{(5)}^{[-3]}$ & $7.130_{(4)}^{[-3]}$ & $1.782_{(3)}^{[-2]}$ & $8.587_{(2)}^{[-2]}$ & $3.393_{(4)}^{[-1]}$ \\
\multicolumn{15}{c}{\vspace*{-2mm}} \\
\hline
\multicolumn{15}{c}{\vspace*{-2mm}} \\
\multirow{1}{*}{${\rm Cu}^{11+}$} &
$4.706_{(2)}^{[+3]}$ & $1.541_{(2)}^{[+4]}$ & $2.982_{(1)}^{[+4]}$ & $7.313_{(2)}^{[+4]}$ & $2.166_{(2)}^{[+5]}$ & $6.000_{(4)}^{[+5]}$ & $1.160_{(1)}^{[+6]}$ &
$5.843_{(3)}^{[-4]}$ & $1.690_{(5)}^{[-3]}$ & $6.440_{(3)}^{[-3]}$ & $1.388_{(1)}^{[-2]}$ & $2.954_{(1)}^{[-2]}$ & $1.494_{(4)}^{[-1]}$ & $3.518_{(2)}^{[-1]}$ \\
\multicolumn{15}{c}{\vspace*{-2mm}} \\
\hline
\multicolumn{15}{c}{\vspace*{-2mm}} \\
\multirow{1}{*}{${\rm Zn}^{12+}$} &
$1.182_{(3)}^{[+3]}$ & $5.335_{(2)}^{[+3]}$ & $1.502_{(1)}^{[+4]}$ & $4.288_{(3)}^{[+4]}$ & $1.626_{(1)}^{[+5]}$ & $4.313_{(1)}^{[+5]}$ & $1.034_{(3)}^{[+6]}$ &
$9.081_{(1)}^{[-5]}$ & $1.299_{(5)}^{[-4]}$ & $1.492_{(5)}^{[-4]}$ & $3.050_{(5)}^{[-3]}$ & $3.209_{(1)}^{[-2]}$ & $1.049_{(1)}^{[-1]}$ & $4.396_{(1)}^{[-1]}$ \\
\multicolumn{15}{c}{\vspace*{-2mm}} \\
\hline
\hline
\end{tabular}
} 
\end{minipage}
\end{table*}

\clearpage
\begin{table}[!htbp]
\begin{minipage}[t]{\textwidth}
\caption{\label{Tab:RR:Maxwl}
Fit coefficients for total ground state RR rate coefficients of recombining ions,
see Eq.~(\ref{Eq:Fit:RR}), where $v_{(u)}^p$ denotes $v(u)\times 10^p$ with uncertainties
given in lower parentheses.
}
\resizebox{\textwidth}{!}{
\begin{tabular}{c l l l l l l}   
\hline
\hline
\multicolumn{7}{c}{\vspace*{-2mm}} \\
$\begin{array}{c} {\rm ion}^{{\rm a}} \\                                       \end{array}$ &
$\begin{array}{c} A                   \\ (10^{-11}\;{\rm cm}^{3} {\rm s}^{-1}) \end{array}$ &
$\begin{array}{c} B                   \\  {\rm  }                              \end{array}$ &
$\begin{array}{c} T_{0}               \\ ({\rm K})                             \end{array}$ &
$\begin{array}{c} T_{1}               \\ ({\rm K})                             \end{array}$ &
$\begin{array}{c} C                   \\  {\rm  }                              \end{array}$ &
$\begin{array}{c} T_{2}               \\ ({\rm K})                             \end{array}$ \\
\hline
\multicolumn{7}{c}{\vspace*{-2mm}} \\
${\rm K}^{+}$   & $4.528^{[+0]}_{(3)}$ & $4.234^{[-1]}_{(2)}$ & $5.931^{[+0]}_{(3)}$ & $2.897^{[+9]}_{(3)}$ & $3.049^{[-1]}_{(4)}$ & $1.645^{[+5]}_{(3)}$ \\
\multicolumn{7}{c}{\vspace*{-2mm}} \\
\cline{2-7}
\multicolumn{7}{c}{\vspace*{-2mm}} \\
${\rm Ca}^{2+}$ & $2.248^{[+1]}_{(4)}$ & $6.605^{[-1]}_{(1)}$ & $6.175^{[+0]}_{(1)}$ & $6.032^{[+6]}_{(1)}$ & $3.158^{[-1]}_{(2)}$ & $2.100^{[+5]}_{(2)}$ \\
\multicolumn{7}{c}{\vspace*{-2mm}} \\
\cline{2-7}
\multicolumn{7}{c}{\vspace*{-2mm}} \\
${\rm Sc}^{3+}$ & $4.196^{[+0]}_{(2)}$ & $2.250^{[-1]}_{(1)}$ & $4.232^{[+2]}_{(1)}$ & $1.110^{[+8]}_{(1)}$ & $3.993^{[-1]}_{(1)}$ & $3.973^{[+5]}_{(3)}$ \\
\multicolumn{7}{c}{\vspace*{-2mm}} \\
\cline{2-7}
\multicolumn{7}{c}{\vspace*{-2mm}} \\
${\rm Ti}^{4+}$ & $3.989^{[+1]}_{(3)}$ & $5.658^{[-1]}_{(2)}$ & $9.517^{[+1]}_{(3)}$ & $4.776^{[+7]}_{(2)}$ & $1.065^{[-1]}_{(4)}$ & $3.534^{[+5]}_{(2)}$ \\
\multicolumn{7}{c}{\vspace*{-2mm}} \\
\cline{2-7}
\multicolumn{7}{c}{\vspace*{-2mm}} \\
${\rm V}^{5+ }$ & $4.193^{[+1]}_{(7)}$ & $5.887^{[-1]}_{(1)}$ & $1.993^{[+2]}_{(6)}$ & $2.207^{[+7]}_{(8)}$ & $9.247^{[-2]}_{(1)}$ & $2.012^{[+5]}_{(7)}$ \\
\multicolumn{7}{c}{\vspace*{-2mm}} \\
\cline{2-7}
\multicolumn{7}{c}{\vspace*{-2mm}} \\
${\rm Cr}^{6+ }$& $5.744^{[+1]}_{(8)}$ & $6.220^{[-1]}_{(1)}$ & $2.282^{[+2]}_{(2)}$ & $2.250^{[+7]}_{(8)}$ & $4.925^{[-2]}_{(1)}$ & $2.948^{[+5]}_{(7)}$ \\
\multicolumn{7}{c}{\vspace*{-2mm}} \\
\cline{2-7}
\multicolumn{7}{c}{\vspace*{-2mm}} \\
${\rm Mn}^{7+ }$& $2.976^{[+1]}_{(5)}$ & $4.838^{[-1]}_{(8)}$ & $1.397^{[+3]}_{(3)}$ & $4.566^{[+7]}_{(8)}$ & $5.863^{[-2]}_{(1)}$ & $1.672^{[+4]}_{(3)}$ \\
\multicolumn{7}{c}{\vspace*{-2mm}} \\
\cline{2-7}
\multicolumn{7}{c}{\vspace*{-2mm}} \\
${\rm Fe}^{8+ }$& $3.341^{[+1]}_{(3)}$ & $4.865^{[-1]}_{(7)}$ & $1.891^{[+3]}_{(7)}$ & $5.181^{[+7]}_{(7)}$ & $5.747^{[-2]}_{(1)}$ & $2.734^{[+4]}_{(9)}$ \\
\multicolumn{7}{c}{\vspace*{-2mm}} \\
\cline{2-7}
\multicolumn{7}{c}{\vspace*{-2mm}} \\
${\rm Co}^{9+ }$& $2.542^{[+1]}_{(3)}$ & $3.600^{[-1]}_{(6)}$ & $5.026^{[+3]}_{(4)}$ & $5.480^{[+7]}_{(1)}$ & $1.142^{[-1]}_{(3)}$ & $2.509^{[+4]}_{(1)}$ \\
\multicolumn{7}{c}{\vspace*{-2mm}} \\
\cline{2-7}
\multicolumn{7}{c}{\vspace*{-2mm}} \\
${\rm Ni}^{10+}$& $3.076^{[+1]}_{(2)}$ & $3.849^{[-1]}_{(1)}$ & $5.284^{[+3]}_{(4)}$ & $4.945^{[+7]}_{(1)}$ & $1.021^{[-1]}_{(1)}$ & $2.571^{[+4]}_{(1)}$ \\
\multicolumn{7}{c}{\vspace*{-2mm}} \\
\cline{2-7}
\multicolumn{7}{c}{\vspace*{-2mm}} \\
${\rm Cu}^{11+}$& $3.527^{[+1]}_{(4)}$ & $4.052^{[-1]}_{(3)}$ & $5.855^{[+3]}_{(1)}$ & $5.691^{[+7]}_{(1)}$ & $8.842^{[-2]}_{(1)}$ & $3.626^{[+4]}_{(1)}$ \\
\multicolumn{7}{c}{\vspace*{-2mm}} \\
\cline{2-7}
\multicolumn{7}{c}{\vspace*{-2mm}} \\
${\rm Zn}^{12+}$& $3.964^{[+1]}_{(2)}$ & $4.107^{[-1]}_{(3)}$ & $6.576^{[+3]}_{(1)}$ & $5.855^{[+7]}_{(1)}$ & $8.595^{[-2]}_{(1)}$ & $3.942^{[+4]}_{(3)}$ \\
\multicolumn{7}{c}{\vspace*{-2mm}} \\
\hline
\hline
\end{tabular}
} 
     ${}^{{\rm a)}}\,$ present work: $\ell\leq 200$, $\;3\leq n\leq 1000$
\end{minipage}
\end{table}

\end{document}